\documentclass[a4paper,12pt]{report}
\usepackage[utf8]{inputenc}
\usepackage{epsfig}
\usepackage{amssymb}
\usepackage{amsmath}
\usepackage{epsfig}
\usepackage{color}
\usepackage{textcomp}

\voffset = - 3cm
\def\beq{\begin{equation}}
\def\eeq{\end{equation}}
\def\ppl{\left(\frac{\phi}{M_4}\right)}
\def\pplc{\left(\frac{\phi_{i}}{M_4}\right)}
\def\pple{\left(\frac{\phi_{end}}{M_4}\right)}
\def\ppls{\left(\frac{\phi_{*}}{M_4}\right)}

\begin{document}
\setcounter{page}{0}
\newpage
\thispagestyle{empty}


\begin{center}
\vfill
{\Large \bf {Inflation in Brane World Gravity}}

\vspace*{1.5cm}

{ \it a  project report submitted for the 
partial fulfillment of the degree
of}\\
\vspace*{2cm}

{\Large{\bf Master of Science}}\\
\vspace*{1.5cm}
{\small \it by}\\
\vspace*{1.5cm}
{\Large \bf Argha Banerjee}\\
\vspace{.5cm}
{\large \bf Roll No. 97} \\
\vspace{.5cm}
{\large \bf Registration No. 13220911001} \\
\vspace*{2cm}

{\large Supervisor: Dr. Ratna Koley}
\vspace*{3.5cm}

\begin{figure}[h]
\begin{center}
\includegraphics[height = 4 cm, width = 6 cm]{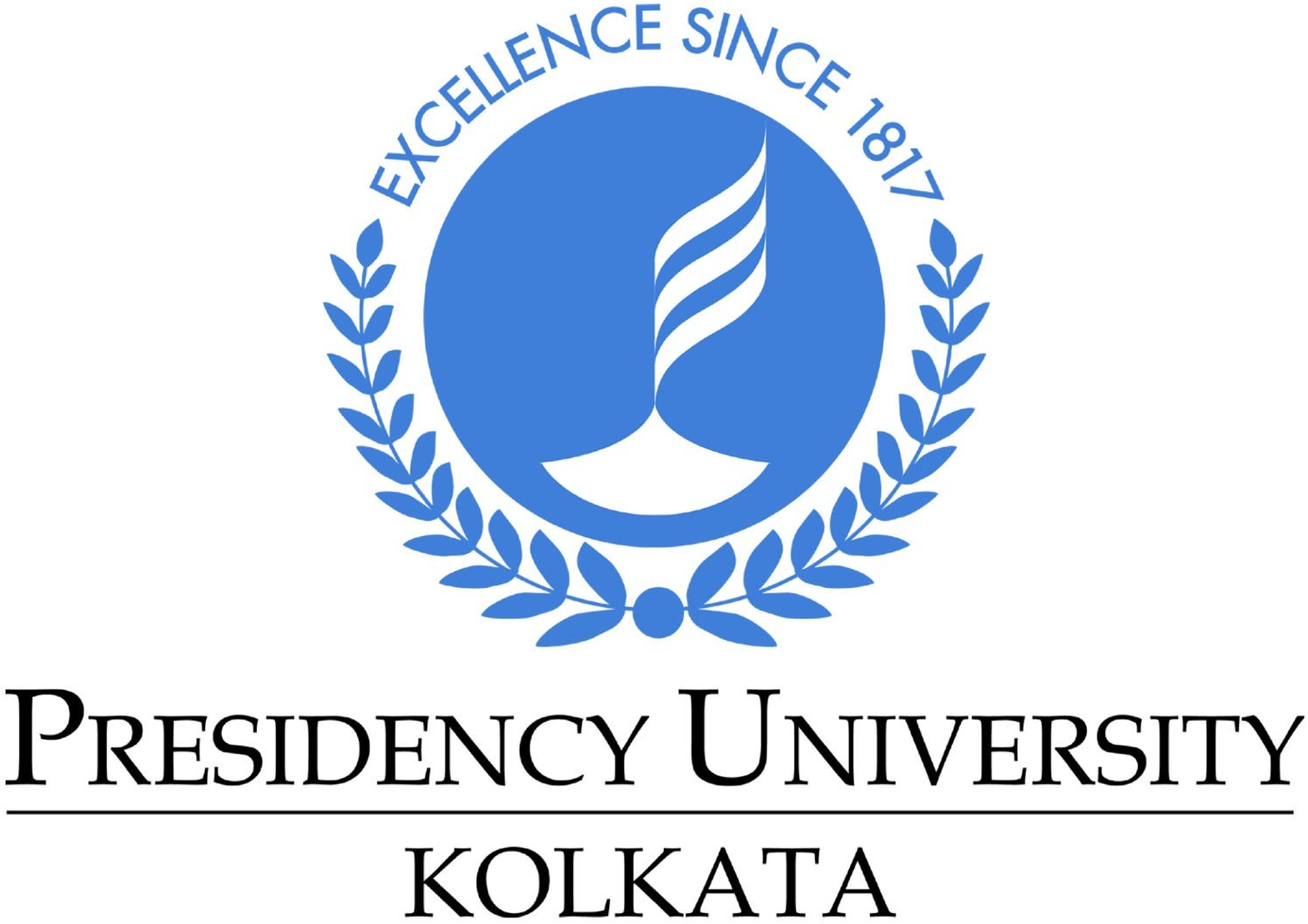}
\end{center}
\end{figure}

{\large Department of Physics \\
86/1, College Street, Kolkata- 700073, West Bengal, India
}

\end{center}
\newpage
\thispagestyle{empty}

\hspace*{3cm}
\begin{figure}
\epsfxsize=1.6in\epsffile{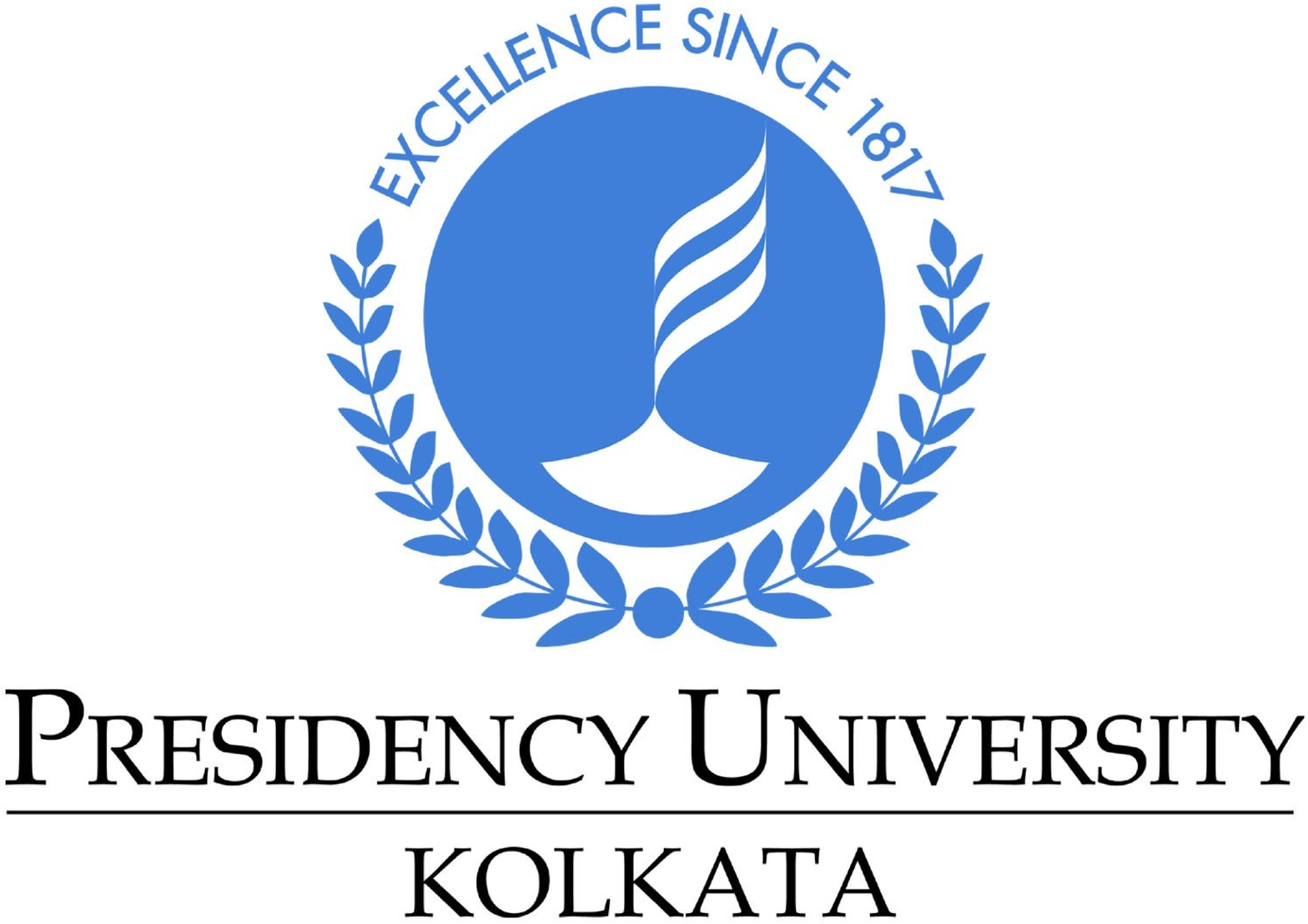}
\end{figure}

\vspace{-4.0cm}
\hspace*{8.7cm} {\Large{\bf Dr. Ratna Koley}}\\

\vspace*{-0.08cm}
\hspace {8.6cm} {\it {\bf Assistant Professor}} \\

\vspace*{-0.08cm}
\hspace*{8.55 cm} {\it {\bf Department of Physics }} \\
\vspace{-0.08cm}
\hspace*{9cm} {\it {\bf Presidency University}}\\
\vspace{-0.08cm}
\hspace*{9cm} {\it {\bf Kolkata, 700073, West Bengal, India}} \\
\vspace{-0.08cm}
\hspace*{9cm} {\it {\bf Email - ratna.physics@presiuniv.ac.in}} \\

\vspace*{3cm}
\begin{center}
{\Large {\it{\bf \underline{Certificate}}}} \\
\end{center}

\vspace{1 cm}
This is to certify that the project report entitled {\bf `Inflation in Brane World Gravity'} which is being submitted to the
Presidency University, Kolkata, in partial fulfillment for the award
of the
degree of Master of Science by {\bf Argha Banerjee} is a bonafide record
of  the research work carried out by him under my supervision.
To my knowledge, the results embodied in this report have not been
submitted to any other university or institute for the award of any degree or
diploma.

\vspace{2.6cm}
\noindent
May 2015
~~~~~~~~~~~~~~~~~~~~~~~~~~~~~~~~~~~~~
~~~~~~~~~~~~~~~~~~~~~~~~~~~~~~~~~~~~~~~~~{\bf Ratna Koley} \\
\noindent Kolkata
~~~~~~~~~~~~~~~~~~~~~~~~~~~~~~~~~~~~~~~~~~~~~~~~~~~~~~~~~~~
~~~~~~~~~~~~~~~~~~~~~( Supervisor )

\newpage

\vspace*{3 cm}
\begin{center}
 {\Large{\bf{Declaration}}}
\end{center}

\vspace*{2 cm}
{\large
I hereby declare that the work reported in this report is
original. It was carried out by me at the Department of Physics of
Presidency University, Kolkata, India, under the
supervision of Dr. Ratna Koley.
The report is based on discovery of new facts and new
interpretation of established facts by others. The author is solely
responsible for  unintentional oversights and errors, if any.
I further declare that it has not formed the basis for the award of any degree,
diploma, membership, associateship or similar title of any other university
or institution.

\vspace {2 cm}

\noindent May 2015\\
Department of Physics \\
Presidency University \\
Kolkata, 700073, India \hspace{7.5 cm} ({\bf Argha Banerjee})}

\newpage

\tableofcontents
\newpage

\begin{abstract}
We study the inflationary dynamics in Brane World gravity and look for observational signatures of any deviation from the standard 
General Relativity based results of Cosmological Perturbation Theory. We first review the standard paradigm of General Relativity 
based inflationary dynamics and cosmological perturbation theory and then go on to review Brane World gravity. Finally we look at 
the high energy corrections for some chosen models and compare the results with the Planck and WMAP (9 year) data. Then we make a 
summary of our results and point out certain interesting features of Brane World gravity based calculations and infer it's implications 
on Brane World gravity itself.
\end{abstract}
\chapter{Introduction}

Big Bang cosmology is the preferred tool to study the evolution and characteristics of our universe.
It accurately describes Hubble's Law, primordial nucleosynthesis, the Cosmic Microwave Background and the features of it's power spectrum. However it fails to describe the
origin of large scale structures in the universe and the inhomogeneities in the CMB (temperature fluctuations). 
These and other conceptual problems led 
Alan Guth \cite{AG} and Andrei Linde \cite{Al} to propose the theory of inflation in the 1980s 
which not only solved the traditional flatness and horizon problems of Big Bang Cosmology 
but also gave an elegant mechanism for explaining the inhomogeneities of the early universe.

\section{Big Bang Cosmology} 

Big Bang Cosmology rests generally on the following assumptions-:

1) The field equations of Einstein's theory of General Relativity hold true.

2) The assumption of the Cosmological Principle i.e. the universe on large scales is homogeneous and isotropic.

3) The assumption of perfect fluid for getting the form of the stress energy tensor.

4) The background space is taken to be the FRW metric which is spatially flat.\\

From these assumptions we get the Friedmann equations, the acceleration equations
and the fluid equations and then
from the equation of state and these equations we can get the evolution of the universe for different components
\cite{Ryden}. 
The FRW metric is given by : 
\beq
ds^2= -dt^2 + a^2(t)\left(\frac{dr^2}{1- kr^2}+ r^2(d\theta^2+\sin^2\theta d\phi^2)\right)
\eeq
where $k$ here is the curvature of the universe.
Due to the isotropy of the universe we can choose the coordinates such that our metric looks like
\beq
\label{metric}
ds^2= - dt^2 + a^2(t) d\chi^2
\eeq

where we $d\chi=\frac{dr}{\sqrt{1-kr^2}}$ and we orient our coordinates such that we only have to consider
radial propagation.

The conformal time is defined as-: $d\tau= \int\frac{dt}{a(t)}$

The advantage of using conformal time is twofold. Firstly it reduces the FRW metric (\ref{metric}) to a flat Minkowski metric multiplied by a the square of the scale factor depending on the conformal time.
$ds^2= a^2(\tau)(-d\tau^2 + d\chi^2)$. 
Secondly if we consider null geodesics, which is the case for light rays, then $ds^2=0$ and we get $
\chi(\tau)=\pm \tau + \rm const.$
\begin{figure}[h!!!!]
\centering
\epsfig{figure=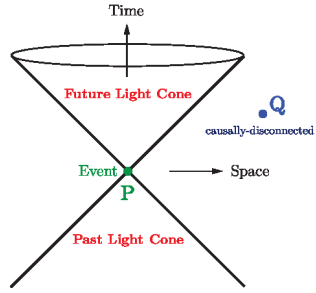,width=4cm,angle=0}
\caption{Figure shows the lightcones in this spacetime diagram.We consider conformal time and thus the lightcones make $45^\circ$
angles in the $\chi-\tau$ plane \cite{Ba1}}
\label{lightcone}
\end{figure}

Thus the spacetime diagram is much easier to construct as the worldlines of the light rays make $45^\circ$ angles in the $\chi-\tau$ plane.
From these considerations we can get an idea about the particle horizon which is defined as the maximum comoving distance that light can travel from an initial time $t_i$ to the present time $t$
due to a signal propagating at the speed of light.
\beq
\chi_{\rm ph}(\tau)=\int_{t_i} ^t \frac{dt}{a(t)}= \tau - \tau_{\rm i}
\eeq
In the conformal spacetime diagram only those spacetime points which are included in the past light cone of the observer have been in
causal contact at any time in the past. This discussions will suffice to point out a glaring anomaly which is found if we
only consider standard Big Bang evolution and which is called the horizon problem.

\subsection{Horizon problem}
\begin{figure}
\centering
\begin{minipage}{.45\columnwidth}
\centering
\epsfig{figure=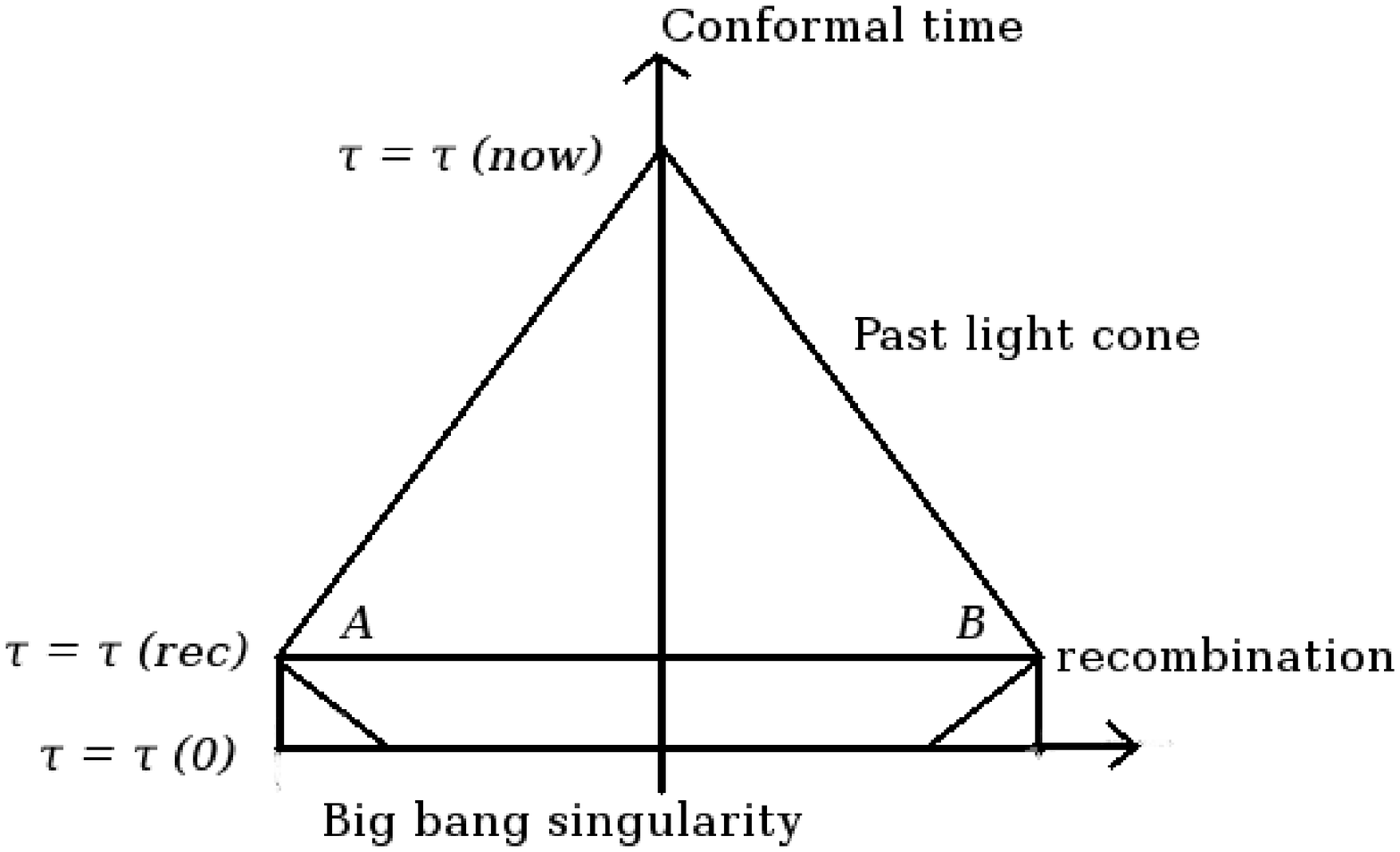,width=6cm,angle=0}
\caption[horizon]{Figure shows two opposite points of the CMB which have never been in causal 
contact as seen from their past light cones.}
\label{horizon1}
\end{minipage}
\hfill
\begin{minipage}{.45\columnwidth}
\centering
\epsfig{figure=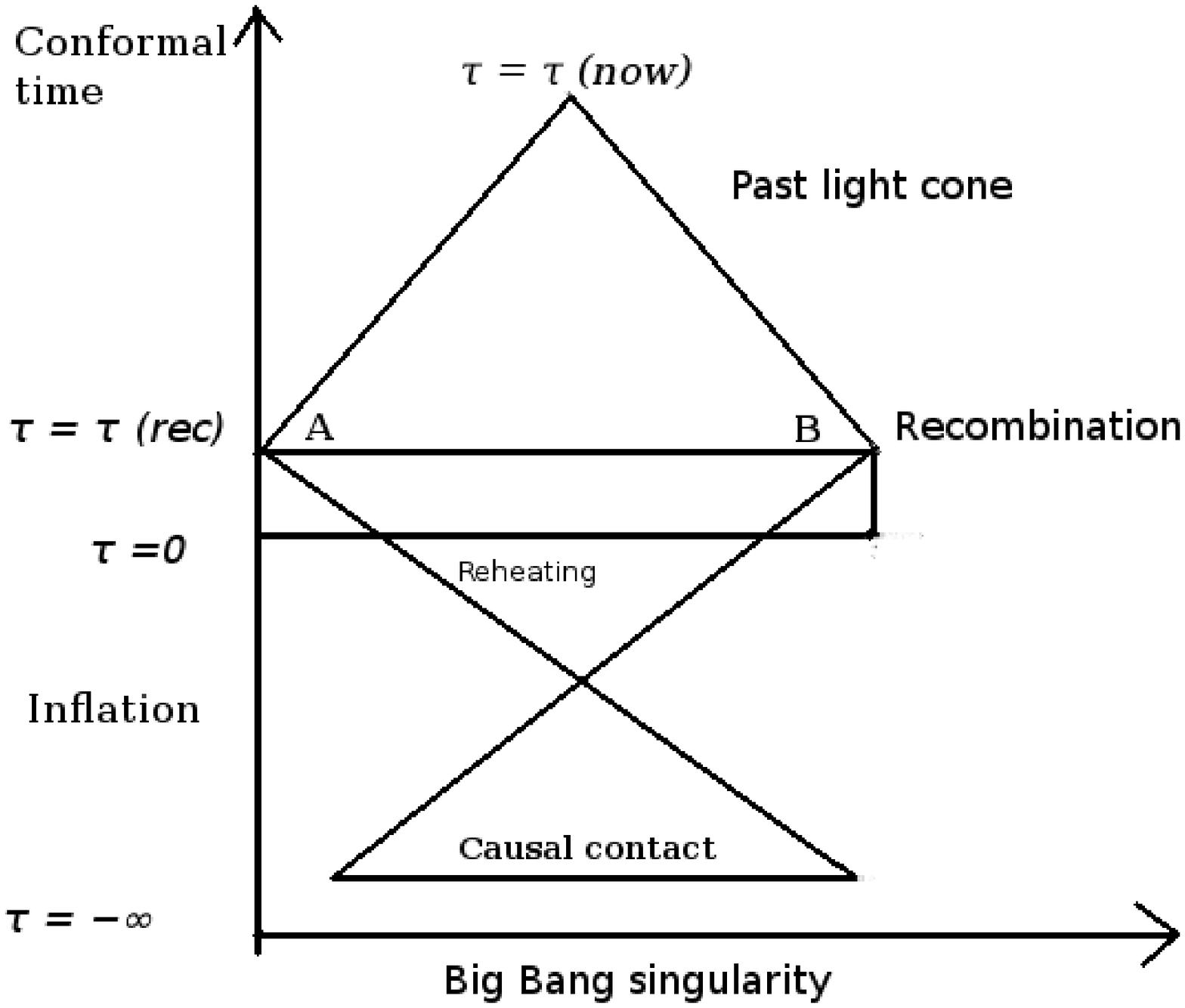,width=6cm,angle=0}
\caption{As the conformal time is pushed back to infinity we can see from the conformal spacetime diagram that the two points of
the CMB which were causally disconnected before are causally connected now.} 
\label{horizon2}
\end{minipage}
\end{figure}

 Consider two opposite points on the CMB (A and B) (Figure \ref{horizon1}) and an observer at time $t=t(now)$. We draw the past light cones of the two opposite points 
 at the time of CMB decoupling. The observer at $t=t_0$ observes a nearly homogeneous and isotropic CMB spectrum as he is
 moving along with the expansion of the universe (i.e, in comoving coordinates). However looking at the
 past light cones of the two CMB points we see that they do not intersect at any point. Thus they could not have been in causal contact
 at the time of decoupling. A detailed analysis paints a far worrying picture as it turns out that points separated by $2^\circ$
 in the CMB sky were causally disconnected at the time of decoupling. This is what is called the horizon problem.
 
Other than the horizon problem we also have the flatness problem and the relic problem that standard Big Bang Cosmology cannot
solve. Our universe is almost flat, however this is an unstable situation and to achieve this degree of flatness Big Bang Cosmology
has to assume extremely fine-tuned initial conditions which leads to the question, what decides this initial conditions? Also
Grand Unified Theories (GUT) predict many relics like magnetic monopoles and topological defects. However we cannot see them now. 
So the question arises that what happened to this unwanted relics?
 \subsection{Need for Inflation}
 
 The problems of Big Bang Cosmology can be seen from the context of the comoving Hubble radius defined as $(aH)^{-1}$ where $H$ is the Hubble parameter
 and $a$ is the scale factor. Physically it is the maximum
 comoving distance that is in causal contact at a particular instant of time. Explicit calculations show that $(aH)^{-1}\propto a^{\frac{1+3\omega}{2}}$
 where $\omega=\frac{p}{\rho}$ and $p$ and $\rho$ are the pressure and density respectively of the dominant component of the universe at any time. However as
 $(1+3\omega)>0$ for the strong energy condition to hold, the Hubble radius always increases in Big Bang cosmology. If we had
 a mechanism of making the Hubble radius decrease in the early universe and then let it increase after a certain time following
 Big Bang evolution then we can get around the problem \cite{Lid}. Scales of cosmological interest (i.e. those scales
 that are inside the horizon now or are entering the horizon just now) would then be inside the horizon at early times
 and would thus be in causal contact. They would then exit the horizon at some time and then re-enter at a later time when standard Big Bang evolution ensues.
  Thus we need a phase of decreasing Hubble radius in the early universe and this was one of the motivations for the theory of inflation.
\begin{figure}[h!!!!]
 \centering\epsfig{figure=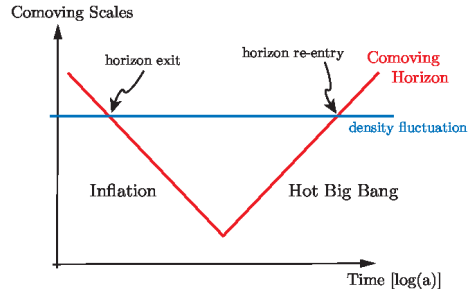,width=6cm,angle=0}
 \caption{The comoving Hubble length decreases during inflation and thus provides a mechanism for explaining the horizon problem.
 Cosmological scales are inside the horizon before inflation starts, they then exit the horizon as the comoving horizon
 decreases during inflation and then re-enters the horizon at a later time after the end of inflation.\cite{Ba1}}
 \label{scales}
\end{figure}

 \section{Inflation}
 
Inflation is a period of accelerated growth of the early universe over a  short period of time \cite{Lid}. The universe underwent exponential growth
during inflation. There are various mathematical realisations of inflation :
\beq
\label{infl}
\frac{d}{dt}{(aH)^{-1}}<0\Rightarrow
\frac{d^2a}{dt^2}>0\Rightarrow
\rho+3p<0
\eeq

The first equation straight away specifies that the comoving Hubble radius decreases.
The second equation shows that inflation is a period of acceleration of the universe and the third is a violation of the Strong
Energy Condition i.e. it gives $\omega<-1/3$ while the Strong Energy Condition requires $\omega>-1/3$. Now if we solve for the conformal time as a function of the scale factor we get-
$\tau\propto\frac{a^{\frac{1+3\omega}{2}}}{1+3\omega}$
where $\omega=\frac{p}{\rho}$.
For $\omega<-1/3$ at $a(t_i)=0$ we get $\tau\rightarrow-\infty$ 

Then we see that the conformal time $\tau$ is pushed back to infinity even as the normal time $t$ is fixed at zero. Thus two points on the CMB can now come in 
causal contact as the horizon itself has been extended to infinity.This shows that the CMB points have ample time to establish
causal contact. If we draw the spacetime diagram again (Figure \ref{horizon2}) we will see that the
two opposite points have indeed come in causal contact at some point in the past. Thus the horizon problem is solved.


\subsection{Physics of Inflation}

The violation of the Strong Energy Condition points out that we cannot satisfy the conditions of inflation by
standard matter or radiation. So we have to look for something else. The simplest solution and one that works very well is that
of a scalar field \cite{Ba1}. In essence we define a scalar field $\phi(x,t)$ and write out the stress energy tensor for it (Equation \ref{stress}). However the symmetries
of our background space help us in making the simplifying assumptions of homogeneity and isotropy and thus the scalar field variable only depends on
time and not position. We call it the inflaton field $\phi(t)$. The field has a potential $V(\phi)$. We have to ensure that under certain conditions
the field will violate the Strong Energy Condition $(\omega<-\frac{1}{3}$) and thus give accelerated expansion (Equation \ref{infl}). 
The stress energy tensor is given as :
\beq
\label{stress}
T_{\mu\nu}=\partial_\mu \phi\partial_\nu\phi-g_{\mu\nu}\left(\frac{1}{2}\partial^\sigma\phi\partial_\sigma\phi + V(\phi)\right)
\eeq
and the action is :
\beq
S_\phi=\int d^4 x \sqrt{-g}\left(-\frac{1}{2}\partial^\mu\phi\partial_\mu\phi- V(\phi)\right)
\eeq
Now invoking the simplifying assumptions of a perfect fluid we get the following forms for the pressure and density : $
p_\phi =\frac{\dot\phi^2}{2} + V(\phi)$ and
$\rho_\phi=\frac{\dot\phi^2}{2} +V(\phi)$

The Friedmann equation and the Klein Gordon equation (which describes the evolution of the $\phi$ field itself) is :
\beq
H^2=\frac{1}{3}\left(\frac{\dot\phi^2}{2}+ V(\phi)\right)
\eeq

and

\beq
\ddot\phi+3H\dot\phi+V(\phi)=0
\eeq

Just from the equations of pressure and density we get for the equation of state parameter $\omega$, 
\beq
\omega=\frac{P}{\rho}=\frac{\frac{\dot\phi^2}{2}+V(\phi)}{\frac{\dot\phi^2}{2}-V(\phi)}
\eeq

Now for $\dot\phi^2<<V(\phi)$ we get $\omega\cong-1$ which satisfies the condition $\omega<-\frac{1}{3}$, i.e. $\rho+3p<0$.
Thus we can see that the scalar field satisfies the conditions for inflation in the regime $\dot\phi^2<<V(\phi)$ which
is called the slow roll approximation.
The slow roll parameters are defined as-  $\epsilon=\frac{\dot H}{H^2}$ and $\eta=\frac{\ddot\phi}{H\dot\phi}$

The parameter $\epsilon$ characterises whether inflation can occur and $\eta$ characterises for how long it will occur.
The slow roll parameters in terms of the potential is given as $
\epsilon_v(\phi)=\frac{M_{pl}^2}{2}\left(\frac{V'}{V}\right)^2$ and 
$\eta_v(\phi)=M_{pl}^2\frac{V''}{V}$
where $M_{pl}$ is the Planck Mass.
 The slow roll conditions are given as $\epsilon_v<<1$ and $\eta_v<<1$ . It is to be noted that these are necessary
but not sufficient conditions for inflation to occur for a period of time. It is seen that $\ddot\phi$ must be small for inflation to
last for a sufficient period of time and the slope of the potential $V(\phi)$ must be flat enough for inflation to occur.
 The Friedmann and the Klein-Gordon equations in the slow roll limit are :
\beq
H^2\approx\frac{V(\phi)}{3}
\eeq
\beq
\dot\phi\approx\frac{-V'}{3H}
\eeq

Inflation stops when the slow roll conditions are violated. Another important quantity that is to be defined is the number
of e-foldings $N$ which characterises how much inflation has occurred. It is defined by the equation :
\beq
N=\ln\frac{a(t_{end})}{a(t_{initial})}=\int_{t_i}^ {t_e} Hdt
\eeq
\beq
\approx\int_{\rm \phi_{end}}^ \phi \frac{V}{V'}d\phi\approx\int_{\rm \phi_{end}}^\phi \frac{d\phi}{\sqrt{2\epsilon_v}}
\eeq

Now to solve the horizon problem we must at least ensure that the cosmological scales entering just now (the observable universe)
were inside the horizon before inflation began and the Hubble radius started decreasing. By calculating the no.of e-folds for this 
scenario it can be explicitly shown that $N\approx 60$ for inflation to solve the horizon problem. It can be seen from the figure
that there is a certain symmetry to the exit and re-entry of cosmological scales. Cosmological scales that enter $N$ e-folds after the
end of inflation had exited the horizon $N$ e-folds before the end of inflation.
%
%
\section{Cosmological Perturbation Theory}
\label{copeth}

The most attractive feature of inflation is that it gives a natural mechanism to seed the inhomogeneities of the CMB and
subsequently the large scale structures seen today \cite{LL1}. Now we cannot have large scale structures or temperature fluctuations from 
a homogeneous and isotropic background. The scalar field variable $\phi(t)$ must now depend on both position and time. The natural candidate for this
are quantum fluctuations $\delta\phi(x,t)$ of the scalar field which depend on both position and time. The different regions of space
will now fluctuate by different amounts. However this will result in the time for the end of inflation to vary for different
regions of space resulting in a time delay $\delta t(x)$. This will subsequently result in fluctuations in the energy density $\delta\rho(x)$
as different regions of space will inflate by different amounts. These density fluctuations will act as the seed for the temperature
fluctuations $\Delta T$ as seen in the CMB.
\beq
\delta\phi(x,t)\rightarrow\delta t(x)\rightarrow\delta\rho(x,t)\rightarrow\Delta T(x)
\eeq

Observations of the CMB tell us that the inhomogeneities are very small (of the order of $10^{-5}$). From that we can understand that the universe
at the time of CMB decoupling was nearly homogeneous with very small inhomogeneities. Thus we can consider linear perturbation theory
and expand the Einstein's equations to only first order 
\beq
\label{Ei}
\delta G_{\rm\mu\nu}=8\pi G\delta T_{\rm\mu\nu}
\eeq

The plan for the rest of the report is as follows. We will first study the theory of linear perturbations and the problem of
gauge choice that comes with it, then we will consider the most general metric perturbations and matter perturbations and
equate them using the perturbed Einstein's equations (Equation\ref{Ei}). Then we will construct gauge-invariant quantities and show why they are
so useful in this context. We will then briefly summarise the problem of vacuum choice in QFT in curved space and then outline
the calculation of the power spectrum in Quasi-de sitter space. Our motivation in all these is to look at how we can relate the
statistical properties of this power spectrum with observable results and we will briefly summarise the results of the Planck
and WMAP experiments and compare the prediction of the existing class of theories with that. This will give us an idea of where
we can probe for signatures of new physics and we will discuss about resolved and open issues pertaining to inflationary dynamics. 
Finally we talk about Brane World gravity as one of the alternatives to the standard G.R. formulation and briefly talk about the
extra features that it predicts and how we plan to probe it's various predictions in the context of inflation.

\subsection{Linear Perturbation Theory}

We can separate any physically observed quantity X as the part of the quantity in the homogeneous background and a perturbation part
that depends on spatial coordinates \cite{BB}.
\beq
\delta x(t,x)= X(t,x)-\bar X(t)
\eeq
where $\delta X<<\bar X$. 
Thus we can split both the matter and metric perturbations as-
\beq
\delta g_{\rm\mu\nu}(t,x)= g_{\rm\mu\nu}(t,x)-\bar g_{\mu\nu}(t)
\eeq
\beq
\delta \phi_{\rm\mu\nu}(t,x)= \phi_{\rm\mu\nu}(t,x)-\bar\phi_{\mu\nu}(t)
\eeq
where the bar overhead in the above equations signifies the value of the quantity in the background spacetime.

However we run into some problems by this definition due to the non-unique way in which we can split any quantity as a background value
and some perturbed part \cite{Ba1} \cite{Dav}. What we are doing necessarily is breaking the perturbation $\delta x$ into a physical spacetime part and a
homogeneous background part. However with this we must also allow for a prescription for the one-to-one mapping between the points
on the two spacetimes. This is called the gauge choice and changing the gauge will result in a change in the perturbations. We can also
create fictitious perturbations and by the same token remove physically relevant perturbations  \cite{BB}. 
For example 
 we can change the time coordinate $\tau$ to $\tau+\sigma^0(t,x)$ and this will result in density perturbations $
\rho(\tau)\rightarrow\rho(\tau+\sigma^0(\tau,x)=\bar\rho(\tau,x)+\bar\rho'\sigma^0$
and the fictitious density perturbation (fictitious as it results from a coordinate change and not due to some physical reason)is
$\delta\rho=\bar\rho'\sigma^0$. The way out of this is to either consider the full set of perturbations in both the matter and
gravitational sector so that we can track all the perturbations or we can use gauge-invariant quantities which by definition do
not change on changing the gauge.
In what follows we will consider gauge-invariant quantities and then choose a particular gauge to simplify the calculations.

\subsection{Metric Perturbation}
\label{metricp}

The most general perturbed metric \cite{Ba1}:
\beq
ds^2=g_{\rm\mu\nu}dx^{\rm\mu}dx^{\rm\nu}=-(1+2\Phi)dt^2+2aB_{\rm i}dx^{\rm i}dt+a^2[(1-2\Psi)\delta_{\rm ij}+E_{\rm ij}]dx^{\rm i}dx^{\rm j}
\eeq

and $B_{\rm i}=\partial_{\rm i}B-S_{\rm i}$ where $\partial_{\rm i}S^{\rm i}=0$
and $E_{\rm ij}=2\partial_{\rm ij}E+2\partial_{(\rm i}F_{\rm j)}+h_{\rm ij}$
and $\partial_{\rm i}F^{\rm i}=0$ and $h_{\rm i}^{\rm i}=\partial^{\rm i}h_{\rm ij}=0$

We will find it easier to work in Fourier space and any quantity $X$ can be fourier transformed by
\beq
X_{\rm k}(t)=\int d^3x X(t,x)e^{i k.x}
\eeq
The various fourier modes denoted by the wavenumber $k$ all evolve independently of each other due to translational invariance
of the background spacetime. We can also separate out the perturbations as scalars,
vectors and tensors. Here $\Phi$, $\Psi$, $B$ and $E$ is the scalar part; $E^{\rm i}$, $F^{\rm i}$ is the vector part and $h_{\rm ij}$ is the tensor part. Vector
perturbations decay rapidly as the universe expands and so we only consider scalar and tensor perturbations. The scalar quantities
all transform with a change of coordinates but the tensor fluctuations don't. These in general facts of Linear Perturbation Theory
can be proved \cite{MW} \cite{Ba1}.

\subsection{Matter Perturbations}

\label{matterp}
In a similar vein we get perturbations of the stress-energy tensor $T_{\rm\mu\nu}$ as-
\beq
\delta T_0^0 =-(\bar\rho+\delta\rho)
\eeq
\beq
\delta T_{\rm i}^0=(\bar\rho+\bar p)v_{\rm i}
\eeq
\beq
\delta T_0^{\rm i}=-(\bar\rho+\bar p)(v^{\rm i}+B^{\rm i})
\eeq
\beq
\delta T_{\rm j}^{\rm i}=\delta_{\rm j}^{\rm i}(\bar p+\delta p)+\Sigma_{\rm j}^{\rm i}
\eeq
where $\rho$ is the density, $p$ the pressure and $\delta q$ is the momentum density and $\Sigma_{\rm j}^{\rm i}$ is the anisotropic
stress.

Now that we have got both the matter (\ref{matterp}) and metric perturbations (\ref{metricp}) we can get the form of the perturbed
Einstein equations (\ref{Ei}) from these \cite{Ba1}\cite{Maar3}. This will help us in finding the time evolution of the gauge-invariant
 quantities that we will consider in {\bf Section \ref{gaugesection}}

\subsection{Gauge Invariant Quantities}
\label{gaugesection}

As already explained we will now construct gauge-invariant quantities. There are various quantities that can be constructed but
we will state two of the physically relevant ones and then explain why they are so important in the context of cosmological perturbations.
They are the curvature perturbation on uniform density hypersurfaces ($\zeta$) and 
the comoving curvature perturbation ($\cal{R}$). The subsequent discussion will explain why we consider these two quantities.
\beq
-\zeta=\Psi+\frac{H}{\dot{\bar\rho}}\delta\rho
\eeq
\beq
{\cal R} =  \Psi-\frac{H}{\bar\rho+\bar p}\delta q=\Psi+\frac{H}{\dot{\bar\phi}}\delta\phi
\eeq

\begin{figure}[h!!!!]
\centering
\epsfig{figure=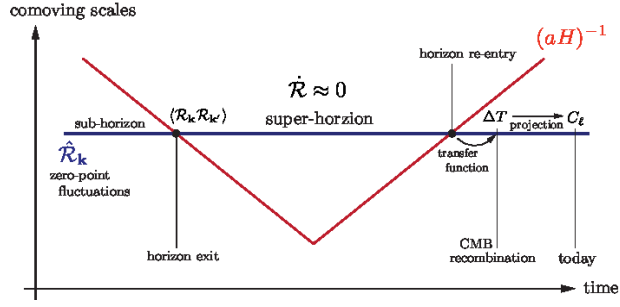,width=8cm,angle=0}
\caption{Comoving cosmological scales exit the horizon as the Hubble radius decreases during inflation and enter again after the 
end of inflation \cite{Ba1}}
\label{fig:sum}
\end{figure}

Now as we can see from Figure \ref{fig:sum} the Hubble radius is decreasing during inflation while the the comoving scales specified by the Fourier modes ($k$) stay constant.
 Thus the modes will exit the horizon at a certain time and re-enter at a certain time after the end of inflation. The evolution of quantities
when they are outside the horizon, defined as super-horizon ($k<<aH$) is uncertain and so we look for gauge-invariant quantities whose
evolution on super-horizon scales is negligible. Explicit calculations \cite{Ba1} show that  on super-horizon scales $\zeta$ and $\cal{R}$ are
equal and from the Einstein equations we can show that that their evolution on super-horizon scales is almost zero
$\dot{\cal R}\approx 0$. Thus if we calculate the value of $\cal {R}$ at horizon exit ($k=aH$) we don't have to worry about super-horizon
evolution and the value at horizon re-entry is the same as at horizon exit. This simplifies the calculations immensely. We will
take advantage of this fact in the ensuing calculations.

\subsection{Power spectrum in Quasi-de Sitter Space}

We will calculate the power spectrum in Quasi-de Sitter space. We make the gauge choice that $\delta\phi=0$ which is
called the comoving gauge choice. We cannot consider complete de-Sitter space as then we won't have any mechanism
to end inflation as we will always have exponential expansion in perfect De-Sitter.
However we will have to end inflation so that we can ensure that standard Big-Bang evolution starts.
 So we consider Quasi-de Sitter space. However as we are working in the framework of Quantum Field Theory in curved spacetime
the vacuum choice is not unique and we have to choose the vacuum considering certain limiting solutions. First we get the action by expanding
the action for scalar perturbation in terms of the fluctuations.
\beq
S=\frac{1}{2}\int d^4x\sqrt{-g}[R-(\nabla\phi)^2-2V(\phi)]
\eeq
\beq
S_{(2)}=\frac{1}{2}\int d\tau d^3x[(v')^2+(\partial_{\rm i}v)^2+\frac{\ddot z}{z}v^2]
\eeq

where $v=z\cal{R}$ is the Mukhanov variable, $R$ is the Ricci scalar and $z=2a^2\epsilon$ where $a$ is the scale factor and
$\epsilon$ is the slow roll parameter. $S_{(2)}$ is the second order action in terms of $\cal{R}$. We can then fourier transform
the Mukhanov variable and get an equation for the mode function $v_k$, the fourier coefficient of the fourier transformed Mukhanov variable ($v$).
Then we get the equation-
\beq
\label{mode}
v_k''+(k^2-\frac{z''}{z})v_ k=0
\eeq
Here the superscript prime denotes differentiation with respect to conformal time. 
 We then follow the standard technique of quantization and promote the mode functions $v_k$ to a quantum operator $\hat v_k$.
\beq
v_k\rightarrow \hat v_k=v_k(\tau)\hat a_k+v_k^*(\tau)\hat a_k^{\dagger}
\eeq
where
\beq
\label{bound}
<v_k,v_k'>=\frac{i}{\hbar}(v_k^*v_k'-v_k^{*'}v_k)=1
\eeq
and
\beq
[\hat a_k,\hat a_k'^{\dagger}]=(2\pi)^3\delta(k-k')
\eeq
where $\hat a_k^{\dagger}$ and $\hat a_k^{\dagger}$ are the creation and annihilation operators respectively. 
The mode functions $v_k$ are specified only by the boundary conditions (\ref{bound}) and they have to satisfy the 
equations of motion (\ref{mode}). However we can have different forms of $v_k$ that can satisfy these two conditions. But the
creation and annihilation operators are defined with respect to the mode functions and any change in the mode functions will
change the creation and annihilation operators. This will subsequently change the vacuum state \cite{Mu} as the vacuum state is defined as-
\beq
\hat a_k\mid 0>=0
\eeq
Thus the vacuum state will change on changing the mode functions and there is no unique choice of the vacuum state. This problem
is common for quantum field theory in curved spacetime.
In what follows we will choose the vacuum state by considering certain limiting conditions.
 We can see that the equation for the mode function (\ref{mode}) is that of a simple harmonic oscillator with an extra time dependent
term added. In the de-Sitter limit ($H$=const) it reduces to-
\beq
\frac{z''}{z}=\frac{2}{\tau ^2}
\eeq
Thus if we take the limit when the mode was deeply inside the horizon and $\tau\rightarrow-\infty$ then we can neglect that term.
Considering this we consider the Bunch-Davies vacuum choice which is taken to satisfy the condition when the mode was well inside the horizon.
\beq
v_k=\frac{e^{-{\rm i} k\tau}}{\sqrt{2k}}\left(1-\frac{\rm i}{k\tau}\right)
\eeq
wher $\tau$ is the conformal time.

\subsection{Connection with observable parameters}

The {\bf{power spectrum}} is defined as the fourier transform of the correlation function for the comoving curvature perturbations
 ${\cal R}$. Mathematically it is given as
$\vartriangle^2_{\cal R}(k)$ -
\beq
\label{p}
<{\cal R}_ k,{\cal R}_k'>=(2\pi)^3\delta( k+ k')P_{\cal R}( k)
\eeq
and
\beq
\label{pvar}
\vartriangle^2_{\cal R}(k)=\frac{k^3}{2{\pi}^2}P_{\cal R}(k)
\eeq
where $P_{\cal R}(\rm k)$ is defined by equations (\ref{p},\ref{pvar})

The correlation function of the ${\cal R}_k$'s are calculated in the chosen vacuum state which for our case is the Bunch-Davies vacuum.
This is why choosing the vacuum state is so important. The scale dependence of the power spectrum, i.e it's dependence on the
wavenumber $k$, is defined by $n_{\rm s}$, {\bf the scalar spectral index}-
\beq
 n_{\rm s}-1=\frac{d\ln\vartriangle^2_{\rm s}}{d lnk}
\eeq
and the {\bf running of the spectral index}, i.e. the dependence of $n_{\rm s}$ on the wavenumber $k$, is given by $\alpha_{\rm s}$-
\beq
\alpha_{\rm s}=\frac{ dn_{\rm s}}{d\ln{\rm k}}
\eeq
Here the subscript {\bf s} signifies that the calculation of the power spectrum is for the scalar fluctuations.
Calculation of the above defined quantities for the Bunch-Davies vacuum and assuming slow roll conditions to hold we get the following
relations for the scalar power spectrum :
\beq
\label{powers}
\vartriangle^2_{{\cal R}}(k)\approx\frac{1}{24\pi^2}\frac{V}{M_{\rm {pl}}^4}\frac{1}{\epsilon_{\rm v}}
\eeq
and for the spectral index
\beq
n_{\rm s}-1\approx2\eta_{\rm v}-6\epsilon_{\rm v}
\eeq
where $\epsilon_{\rm v}$ and $\eta_{\rm v}$ are the slow roll parameters and both the power spectrum and the spectral index are
calculated at horizon exit $k = aH$  as the evolution of of $\cal{R}$ is approximately equal to zero 
at superhorizon scales, i.e. $\dot{\cal{R}}\approx0$ (\ref{gaugesection})

A similar calculation can be done for the tensor perturbations \cite{Ba1}. However it is seen that the action for the tensor perturbations 
are actually just two copies of the scalar perturbations,one for each polarisation mode. Thus we can straight away write
the power spectrum for the tensor fluctuations with the help of the calculations done for the scalar perturbation. 
That comes out to be for the power spectrum of the tensor fluctuations $\vartriangle_t(\rm k)$ as :
\beq
\label{powert}
\vartriangle^2_{\rm t}(k)\approx\frac{2}{3\pi^2}\frac{V}{M_{\rm {pl}}^4}
\eeq
and the spectral index of the tensor perturbations $n_{\rm t}$ as :
\beq
\label{nt}
n_{\rm t}\approx-2\epsilon_{\rm v}
\eeq
Again this calculation is done for horizon exit $k=aH$.\\

We also define a tensor-to-scalar ratio to quantify the amplitude of the tensor signal 
w.r.t. to the scalar signal as the tensor contribution is very weak compared to the scalar contribution. It is defined as :
\beq
r=\frac{\vartriangle^2_{\rm t}(k)}{\vartriangle^2_{{\cal R}}(k)}
\eeq
In terms of the slow roll parameters it is given as :
\beq
\label{ratio}
r=16\epsilon_{\rm v}
\eeq
From eqns (\ref{nt}, \ref{ratio}) we get the following consistency relation :
\beq
\label{con}
r=-8n_{\rm t}
\eeq
Thus if we can measure the tensor-to-scalar ratio and the spectral index of tensor perturbation independently then it can act
as a powerful test for the theory of single field slow roll inflation.\\

We can also define a quantity called the Lyth bound as :
\beq
\frac{\vartriangle\phi}{M_{\rm{pl}}}\approx\left(\frac{r(N_{\rm CMB})}{0.01}\right)
\eeq
For $r>0.01$ we have $\vartriangle\phi>M_{\rm pl}$ and we have large field inflation.\\

The scalar spectral is is significant because any deviation from 1 will tell us that the power spectrum is slightly
dependent on $k$ and that the Harrison-Zeldovich case for $n_{\rm s}=1$ is not exactly true. The tensor-to-scalar ratio is significant
as  any exact experimental result of it (as opposed to upper bounds as is available from experiments now)
 will confirm the existence of primordial gravity waves.
 
The results of the Planck \cite{Pl} and WMAP \cite{Wm} are given in tabular form \\

\begin{tabular}{|c|c|c|}
 \hline
 Observational Parameters & Planck & WMAP (2012)\\
 \hline
 $r$(Tensor-to-scalar ratio) & $<0.11$ & $<0.13$\\
 \hline
 $n_s$(scalar spectral index) & $0.9616\pm0.0094$ & $0.972\pm0.013$\\
 \hline
 $\vartriangle_s^2$ & $(2.23\pm0.16)\times10^{-9}$ & $(2.41\pm0.10)\times10^{-9}$\\
 \hline
 $\alpha_s$ & $(-0.013\pm0.009)$ & $(-0.019\pm0.025)$\\
 \hline
\end{tabular}\\\\


\section{Various models of Inflation}
\label{modinf}

From  recent Planck results we get $\vartriangle_{\rm s}^2 \approx2.23\times10^{-9}$ and from this we can get an upper bound on the energy scale of
inflation from the equations (\ref{powers})(\ref{ratio}) \cite{Lyth}\cite{Lid}
\beq
\label{energy}
V^{1/4}\approx10^{16}GeV
\eeq

This is again a very significant result as we want to know exactly when inflation occurred in the early universe, i.e.
at what energy scale it took place. More precise experiments will help us in this quest. A very significant detection in 
this regard would be detection of the power spectrum of tensor perturbations $\vartriangle_t^2$ (\ref{powert}) as it depends 
explicitly on the potential $V$ which gives the energy scale of inflation. However it is to be noted that the result of equation
(\ref{energy}) is in the range of what is predicted by GUT theories \cite{Lid}.

There are different classes of inflationary models \cite{Lid}, \cite{Dav},\cite{martin},\cite{ring} : \\

\noindent 1) {\bf {Large Field Models }}: This models are characterised by the condition that the field moves over distances comparable to or greater
than the Planck Mass, i.e. $\vartriangle\phi=M_4$ and it is displaced from it's stable minimum by that amount.
. The slow roll parameters are given by the following conditions $0<\eta\leq\epsilon$. This class of models include the chaotic
inflation models \cite{Lind1} which are one of the most researched models in the literature. In chaotic models
any potential field can cause inflation provided it has a sufficiently flat slope such that slow roll conditions
 can hold. They include-:

$a$)Polynomial inflation \cite{Lid} - $V(\phi)=\lambda\phi^p$

$b$)Power-law inflation \cite{Dav} - $V(\phi)= V_0\exp(\sqrt{\frac{16\pi}{\rm p}}\frac{\phi}{m_{\rm Pl}})$
 
 where $\lambda$, $p$ are parameters of that particular class of theories. For example $p$ can be 1,2,3.... for different
 polynomial potentials that are considered.\\
 
\noindent 2) {\bf Small field models} - This models are characterised by the condition that the the scalar field rolls away from an unstable maximum
of the potential towards a minima of the potential and $\vartriangle\phi<M_{4}$. The slow roll parameters are given by
 $\eta<0<\epsilon$. It is given by the potential form \cite{Dav}- $V(\phi)=\lambda^4\left[1-\left(\frac{\phi}{\mu}\right)^p\right]$
 
 The recent Planck resutls favour the small field models while slightly disfavouring the chaotic inflationary models. There have
 been a number of analyses done on favourable inflationary models \cite{ring} and one of the models that have passed the cutoff
 is the Hilltop inflationary Model \cite{Lyth}. The form of the inflationary potential is- $V(\phi)=V_0-\frac{1}{2}{\rm m}^2\phi^2+....$ .
 The potential has a maximum at $\phi=0$ and the field rolls down from the maximum.
  This models have a negligible tensor to scalar ratio-
 \beq
 r<0.002\left(\frac{60}{N}\right)^2\left(\frac{\phi_{\rm end}}{M_{4}}\right)
 \eeq
 and as $\phi<<M_{4}$ we can see why we get a small $r$. $N$ here is the number of e-folds.There are different forms of the 
 hilltop potential depending on which term of the power series dominates. After Planck 2013 it is seen that only the model
 where the power of $\phi$ in the potential is 2 survives the bounds given by the Planck results. The cases where the power is 3
 or 4 is disfavoured by the Planck results.
\section{Key Predictions of Inflation}

There are some key predictions of inflation :

1) The universe is flat or is very close to flatness. In a way inflation sets the fine-tuned initial condition that we have to assume
in Big Bang Cosmology. 

2) The perturbations produced by inflation are adiabatic and gaussian. However recent experiments have given values of $f_{\rm {NL}}^{\rm local}=2.7\pm5.8$.
This parameter describes the local non-Gaussianity, i.e. it modifies the expected Gaussian estimate for a physical entity. However presence of non-Gaussianity of O(1) is expected in almost all inflationary theories.

3) The spectral index $n_{\rm s}$ is very close to 1 and inflation predicts an almost flat spectrum. However the deviation from 1
is very significant and is an important prediction of inflation as it rules out the Harrison-Zeldovich spectrum which predicts that $n_{\rm s}$
is exactly equal to 1.

4) The CMB spectrum has a number of peaks at different values of $l$, the multipole moment. The peaks can only be explained if the fluctuations produced for every
wavenumber $k$ had the same phase so that they could interfere to form those peaks \cite{Dod}. Inflation predicts fluctuations which have the same
phase for any wavenumber $k$ and thus emerges as a natural candidate to explain the formation of the peaks seen in the CMB spectrum.

5) Inflation produces  scalar, vector and tensor perturbations. The scalar perturbations lead to the density perturbations,
the tensor perturbations are the source of gravitational waves and the vector perturbations decay and are not considered in this report.

\section{Research Proposal}

We have described the theory of inflation, elucidated how it plays a role in providing the seed for structure formation and discussed
it's predictions. However the subject is not a closed book and there exists avenues worth exploring both in terms of future tests
of inflation and for exploring various aspects of the theory of inflation.
There is also ample scope for further work :
 
1) We have so far described the inflationary dynamics assuming that Einstein's formulation of General Relativity hold true
at those energy scales. However we cannot be confident enough of that prediction as we do not have any direct probe into
that energy scale to test our ideas. The data we have in our hand is from studying the CMB and Large Scale Structures. It is generally
expected that some low energy realisation of an yet unknown theory of quantum gravity would hold in these scales. However it
isn't necessary that the low energy realisation be Einstein's G.R., it could be some other theory of gravity and work
can be done to probe what additional effects we will see with this new theory and look for experimental signatures of
such corrections. In fact inflationary dynamics can be used to rule out existing mode gravity theories.

2) We have assumed that inflation arises due to the presence of a scalar field. But could inflation arise from some entity other
than a scalar field component in $T{\mu\nu}$ or could it even be a projection of some higher-dimensional effect?

3) The recent Planck data shows the possibility of a blue tilt,i.e a positive tensor spectral index of the tensor power spectrum.
However most models predict $n_t<1$. Thus this can be an avenue worth exploring.

In that context we consider {\bf Brane World gravity}. Brane World gravity is a theory of gravity where we consider a (1+3+d) dimensional spacetime with a (1+3) dimensional brane 
embedded in a (1+3+d) dimensional bulk \cite{Maar1}. The standard particle fields and particles reside on the brane while gravity can leak into the bulk.
However at low energies gravity is confined on the brane. This formalism results in the addition of several correction terms in
the standard Friedmann equations. We will work with simplest models of this type with $d=1$. Thus we will be considering
a 5 dimensional spacetime with a (1+3) brane embedded in a 5 dimensional bulk.
 Considering the corrections in the high energy regime as given by Brane World gravity we plan on
investigating the various statistical parameters of cosmological perturbations 
(done in the frame work of Brane World gravity \cite{Maar2}) such as the power spectrum, the spectral index 
and the tensor-to-scalar ratio and then comparing the values obtained with available experimental results such as Plank and
WMAP. This could rule out models of the scalar potential $V(\phi)$ that satisfy the conditions for Einstein's G.R. or rule in
other models that had previously over-stepped the cut-off limit. This will help when in future higher-precision experiments will give
better constrained values of the tensor-to scalar ratio $r$ and could act as a falsifiability check of brane world gravity
in the context of inflation. We want to specifically look for deviations from the standard paradigm of General relativity based
perturbation calculations.

\chapter{Brane World Cosmology}
\label{braneworld}
\section{Historical Prelude}

The first theory of extra dimensions was given by Nordstrom \cite{no}. Subsequent work done by Kaluza-Klein \cite{kk} acted as a precursor to later works of 
unifying the different forces by increasing the dimensions of the space-time. Kaluza and Klein were looking 
to unify electromagnetism and gravity by their theory and their approach spawned a generation of such attempts to solve 
physical problems by postulating extra dimensions. One major application of it was in Quantum Gravity. 
While the other forces had been unified, unifying gravity with the other forces remained 
a formidable challenge and one theory that came to prominence in the 1970-80's was String Theory \cite{str}. String Theory 
assumed the universe to be a 10 dimensional entity with 9 spatial dimensions and 1 temporal dimension. The main entities 
of this theory are strings whose oscillations, it was postulated, gives rise to all the interactions that we can see today. 
Open strings were postulated to represent the matter sector, i.e. the Standard Model particles while closed strings 
represented graviton modes thus incorporating gravity. 
However the theory soon ran into problems as it transpired that there was not one but five disparate versions of 
String theory that could be derived. This was solved in the 1990's by the introduction of the M-theory \cite{m} which assumed 
that the Universe had 11 dimensions with 10 spatial and 1 time dimension. The various forms of the 10 dimensional 
string theories are just different low energy realisations of this general theory. 

However to get experimentally 
testable predictions physicists had to look at low energy realisations of String theory and one very promising sector 
was Brane World Gravity \cite{br} \cite{br1} \cite{br2} \cite{br3} \cite{br4} \cite{Maar1}.
 
The word brane is derived from the word membrane and in the literature it usually signifies the subspace of a 
higher dimensional spacetime which is called the bulk. These are just technical terms to differentiate between the 
spacetime slices and the general spacetime. In this and subsequent chapters of this report we will work exclusively 
in the domain of Brane World Gravity.

\section{Brane World Gravity: An Introduction}
\label{BG1}

In Brane World Gravity we consider a (1+3+d) dimensional spacetime with an (1+3) brane embedded in an (1+3+d) dimensional bulk.
The Standard Model fields and particles are confined on the brane while gravity  can access the bulk at high energies. 
However at low energies gravity is also localised on the brane. There exists many scenarios by which this confinement 
of gravity on the brane at low energies can be brought about. The one which we will consider in this chapter is the 
Randall-Sundrum(RS) Models \cite{rs1} \cite{rs2}. In the RS models we consider a curved bulk which is an Anti-DeSitter spacetime and thus has 
a negative cosmological constant. The curvature of the bulk is what forces the gravitational field to be localised on 
the brane at low energies. The brane also has a brane tension which is postulated to negate the effects of the negative 
cosmological constant of the bulk and also provide a mechanism to get self-gravity of the brane. We also consider that 
the  brane has mirror symmetry called $Z_2$ symmetry. The $Z_2$ symmetry is the identification of points with the prescription 
x $\leftrightarrow$ -x. Physically it means that if moving along the extra dimension we moved in from one side passed the 
brane and emerged 
from the other side, the bulk spacetime would look the same on both sides.
\begin{figure}[h!!!!]
\centering
\epsfig{figure=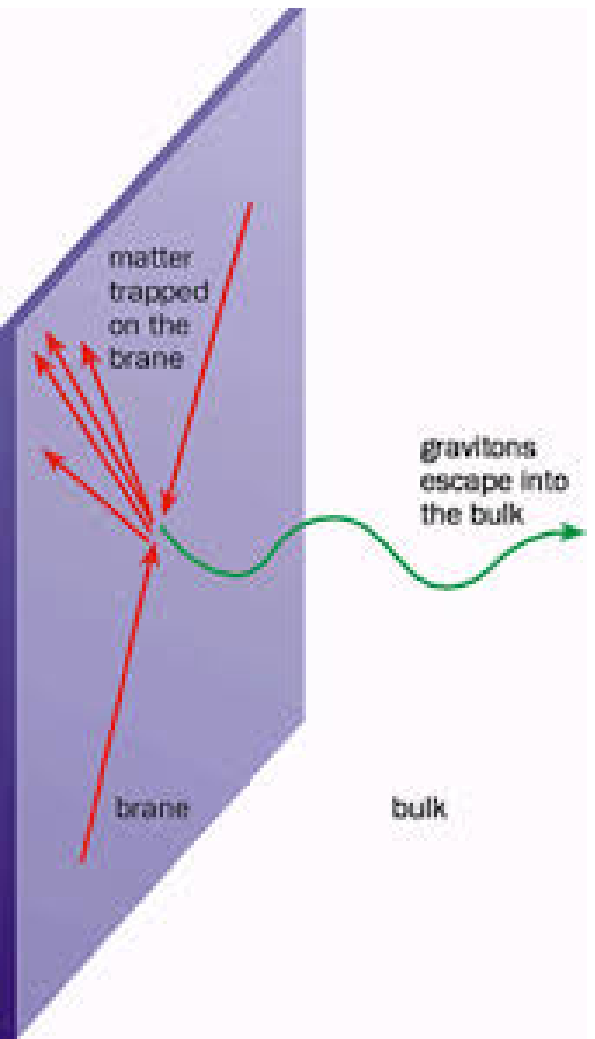,width=7cm,height=8cm,angle=0}
\caption{Brane and Bulk \cite{Maar1}}
\end{figure}

In what follows we will take d=1 and will thus consider a 4 dimensional brane embedded in a 5 dimensional bulk. 
The Einstein equations of G.R.(General Relativity) is expected to be substantially changed due to extra dimensional effects 
and the field equations relevant to Brane World gravity is called the effective field equations on the brane.
 Our motivation here is to look for corrections to the G.R. based calculations and then estimating the source of this
 corrections and the energy scales at which they will be dominant. We will then proceed to show how the Friedmann equations,
 the equation of state and the sound speed are changed due to the effect of this corrections to the field equations. 
 We will then 
move to what is our primary aim and that is to look for the changes in the parameters of the cosmological perturbation analysis 
on Brane World gravity. This will give us the tools to calculate the values of the cosmological parameters for different scalar
field potentials with potential V($\phi)$ and then make certain observations of the results that we get.
  
\section{Effective Field Equations on the Brane}

We will consider the Shiromizu-Maeda-Sasaki approach \cite{sms} to get the Effective Field Equations. Basically
we will get the 4-D(Dimension) brane effects by projection of the 5-D curvature on the brane. We will assume that the Einstein equations
hold in 5 dimensions i.e. in the bulk.
\beq
\label{5Df}
{}^{(5)}G_{ AB}= -\Lambda_5 \; {}^{(5)}g_{AB} + \kappa^2_5 \; {}^{(5)}T_{AB}
\eeq

where ${}^{(5)}G_{AB}, {}^{(5)}g_{AB}$ and ${}^{(5)}T_{AB}$ are the 5-dimensional Einstein tensor, the metric and the energy
momentum tensor due to fields in the bulk respectively. $\Lambda_5$ is the 5-D cosmological constant and $\kappa^2_5$ is the 5-D
 gravitational constant which is related to the 5-D Planck mass $M_5$ by $\kappa^2_5=\frac{8\pi G}{M^3_5}$ and $A,B$ run from
 0 to 4.
 
The 5-D line element is given by-
\beq
{}^{(5)}ds^2= g_{\mu\nu}(x^{\alpha},y) dx^{\mu}dx^{\nu} + dy^2
\eeq

where $g_{\mu\nu}$ is the induced metric on the brane (y= const. surface) and $y$ characterises the extra dimension. Then if we
consider a unit normal to the brane $n^A$, we get $n_A dX^A= dy$. The 5-D metric can then be written as-
\beq
{}^{(5)}g_{AB}= g_{AB} + n_A n_B
\eeq

which can be understood as a decomposition into a 4-D metric $g_{AB}$ and a component of the 4-D metric along the normal to the 
brane. This completely specifies the 5-D metric in terms of the 4-D metric. We will now project the 5-D curvature to the 4-D brane
to get the 4-D curvature.

To do that we use the Gauss-Codazzi equations \cite{gaco}. The Gauss equation projects the 5-D curvature as a 4-D curvature term with
extrinsic curvature terms added on. The extrinsic curvature specifies how the brane (y=const. surfaces) are embedded in the 
general spacetime, i.e the bulk. So the 5-D curvature is decomposed into a 4-D curvature plus the corrections due to how the brane
is itself embedded in the bulk. The Gauss equation can be written as-
\beq
\label{gauss}
R_{ABCD}={}^{(5)}R_{EFGH}\; g_A{}^E g_B{}^F g_C{}^G g_D{}^H + 2 K_{A[{}_{C}K_D]B}
\eeq

where $K_{AB}$ is the extrinsic curvature of the of the y=constant surfaces and the square brackets [ ] signifies antisymmetrization
. Mathematically the extrinsic curvature has the following
 property.
\beq
\label{ec}
K_{AB}= g_A{}^C\; {}^{(5)}\nabla_{C} n_B
\eeq
where $\nabla_C$ is the covariant derivative and $K_{AB} n^B=0$

The Codazzi equation relates the variation of the extrinsic curvature $K_{AB}$ along the brane (y= const.) and is given by :
\beq
\label{coda}
\nabla_B K_A{}^B - \nabla_A K= {}^{(5)}R_{BC} \; g_A{}^B n^C
\eeq
where $K$ is given by $K_A{}^A= K$
 
Using the Gauss-Codazzi equations (\ref{gauss}, \ref{coda}) and the 5-D field equation (\ref{5Df}) we get :
\begin{eqnarray}
\label{4Df}
G_{\mu\nu}= -\frac{1}{2}\Lambda_5 g_{\mu\nu} + \frac{2}{3} \kappa_5^2\;[\;{}^{(5)}T_{AB}\;g_{\mu}{}^A g_{\nu}{}^B +
(\;{}^{(5)}T_{AB}\; n^A n^B - \frac{1}{4}\; {}^{(5)}T)\; g_{\mu\nu}] \nonumber \\
+ K K_{\mu\nu}- K_{\mu}{}^{\alpha}K_{\alpha\nu}+ \frac
{g_{\mu\nu}}{2}\;[K^{\alpha\beta}K_{\alpha\beta}-K^2]\;g_{\mu\nu}-{\cal E}_{\mu\nu}
\end{eqnarray}
where ${\cal E}_{\mu\nu}$ is the projection of the 5-D (bulk) Weyl tensor ${}^{(5)}C_{ABCD}$ on the brane and the form of
 the ${\cal E}_{\mu\nu}$ is given as :
 \beq
{\cal E}_{\mu\nu}= {}^{(5)}C^{ACBD}\; n^C n^D g_{\mu}{}^A g_{\nu}{}^B
\eeq

and ${}^{(5)}T= {}^{(5)}T_A{}^A$

This is the form of the field equations in the bulk coordinates ($X^A$). However what we want is to express the field equations 
completely in the brane coordinates $(x^{\mu})$ and find the form of the field equations on the brane. It can be achieved by
 letting $y\rightarrow \pm 0$ in the field equations. To do this we first consider the total Energy-Momentum tensor
on the brane as
\beq
T_{\mu\nu}{}^{brane}= T_{\mu\nu}- \lambda g_{\mu\nu}
\eeq
where $T_{\mu\nu}$ is the E-M tensor for particles on the brane and $\lambda$ is the brane tension.

The 5-D field equations(\ref{5Df}) with a contribution from the brane is given as
\beq
\label{5Df1}
{}^{(5)}G_{AB}= - \Lambda_5 {}^{(5)}g_{AB}+ \kappa_5^2\;[{}^{(5)}T_{AB}+ T_{AB}{}^{\rm{brane}} \delta(y)]
\eeq
where the delta function is the mathematical realisation of the argument that the standard model particles and fields are 
confined on the brane. From (\ref{5Df1}) by taking the limiting value on the brane we get the Israel-Darmois junction conditions
 which are given as
\beq
\label{ID1}
g_{\mu\nu}^+ - g_{\mu\nu}^- = 0
\eeq
\beq
\label{ID2}
K_{\mu\nu}^+ - K_{\mu\nu}^-= -\kappa_5^2\;[T_{\mu\nu}{}^{\rm{brane}}- \frac{1}{3} T^{\rm{brane}} g_{\mu\nu}]
\eeq

where $T^{\rm{brane}}= g^{\mu\nu} T_{\mu\nu}{}^{\rm{brane}}$. Equation (\ref{ID1}) can be interpreted as a continuity equation 
for the metric coefficients and equation (\ref{ID2}) shows the change in the extrinsic curvature term on passing the brane
 from one side to the other along the extra dimension.
 
We have already seen in section (\ref{BG1}) that we consider the brane to have $Z_2$ symmetry i.e. it is mirror symmetric.
What changes as we move in from one side, cross the brane and emerge on the other side is that the sign  of the normal changes 
for the two sides, it actually reverses. However from equation (\ref{ec}) we can see that this condition requires that the 
extrinsic curvature has the following property $K_{\mu\nu}{}^+=-K_{\mu\nu}{}^-$ as the direction of the normal reverses. Using 
this property and the junction conditions (\ref{ID1}, \ref{ID2}) we have,
\beq
\label{ec2}
K_{\mu\nu}=-\frac{1}{2}\kappa_5^2\;[T_{\mu\nu} + \frac{1}{3}(\lambda- T)\; g_{\mu\nu}]
\eeq
where $T=T_{\mu}{}^{\nu}$

Now using equations(\ref{4Df}, \ref{ec2}) we have finally the effective field equations on the brane
\beq
\label{efe1}
\boxed{G_{\mu\nu}=-\Lambda g_{\mu\nu} + \kappa^2 T_{\mu\nu} + \frac{6\kappa^2}{\lambda}S_{\mu\nu}-{\cal E}_{\mu\nu}+ 4\frac{\kappa^2}{\lambda}
F_{\mu\nu}}
\eeq
where $\Lambda=\frac{1}{2}(\Lambda_5+\kappa^2\lambda)$ is the 4-D cosmological constant and 
\beq
\label{k2}
\kappa^2=\frac{1}{6}\lambda\kappa_5^4
\eeq
 is the 4-D gravitational constant.

\subsection{Correction Terms in Brane-World Gravity}

We will now list the correction terms in the field equations, the important physical parameters, the slow roll parameters and the 
cosmological perturbation parameters. These will help to clearly identify and differentiate between the correction terms in 
the various quantities while maintaining the unifying theme that all these corrections stem from the correction 
terms in equation (\ref{efe1}).

\subsubsection{Correction to the Field Equations}

We can see from equation(\ref{efe1}) that there are a number of correction terms in Brane World gravity when compared to 
standard G.R. based field equations. The correction term $F_{\mu\nu}$ contains ${}^{(5)}T_{AB}$ terms which is the 
E-M tensor for fields on the bulk
(except the cosmological constant).
\beq
\label{f}
F_{\mu\nu}={}^{(5)}T_{AB}\; g_{\mu}{}^A g{\nu}{}^B + [\;{}^{(5)}T_{AB}\;n^A n^B-\frac{1}{4}{}^{(5)}T]\;g_{\mu\nu}
\eeq
We get the analogue of the conservation equation in Brane World gravity from equations (\ref{5Df},\ref{coda},\ref{ec2})
\beq
\label{conb}
\nabla^{\nu} T_{\mu\nu}=-2{}^{(5)}T_{AB}\;n^A g_{\mu}{}^B
\eeq
whereas in Einstein's G.R. we get
\beq
\label{conb1}
\nabla^{\nu}T_{\mu\nu}=0
\eeq

This shows that in Brane World gravity there is inherently the possibility of exchange of matter between the brane and the bulk
 from equation (\ref{conb}). However in our work we will assume that ${}^{(5)}T_{AB}$ is zero (empty bulk) and thus from equation (\ref{f})
  we can see that the correction term $F_{\mu\nu}$ is also zero and thus there is no exchange of matter between the brane and 
  the bulk. We will see that the imposition of this condition is of great help when we will consider perturbative analysis in 
  Brane World gravity. So even in Brane World Gravity the conservation equation reduces to Equation (\ref{conb1})

 The next correction term from equation(\ref{efe1}) is $S_{\mu\nu}$ which is given as
\beq
S_{\mu\nu}=\frac{1}{12} T T_{\mu\nu}-\frac{1}{4} T_{\mu\alpha}T_{\nu}{}^{\alpha}+\frac{1}{24}g_{\mu\nu}
[3T_{\alpha\beta}T^{\alpha\beta}-T^2]
\eeq
$S_{\mu\nu}$ contains terms quadratic in the E-M tensor and thus this is an extra matter term in the existing equations. However 
this term is dominant only at high energies ($\rho>>\lambda$) where $\rho$ is the density. The $S_{\mu\nu}$ term gives the 
corrections to the field equations due to local effects on the brane.

The next correction term is the ${\cal E}_{\mu\nu}$ term which is the projection of the 5 D Weyl tensor on the brane. This does 
not depend on the brane and it is the non-local effects due to the presence of the bulk. 
This has several interesting consequences as we will see 
in the following sections.

Thus finally the field equations from the brane and bulk are from equations (\ref{5Df}, \ref{efe1})-\\

For the bulk
\beq
{}^{(5)}G_{AB}=-\Lambda_5\;{}^{5}g_{AB}
\eeq
For the brane
\beq
\label{brane}
G_{\mu\nu}=-\Lambda g_{\mu\nu}+ \kappa^2 T_{\mu\nu}+ 6 \frac{\kappa^2}{\lambda} S_{\mu\nu}-{\cal E}_{\mu\nu}
\eeq

as we have already assumed that $F_{\mu\nu}$ and ${}^{(5)}T_{AB}$ are zero.

From the Bianchi identities $\nabla^{\nu}G_{\mu\nu}=0$ and equation(\ref{brane}) we get
\beq
\nabla^{\mu}{\cal E_{\mu\nu}}=\frac{6\kappa^2}{\lambda} \nabla^{\mu}S_{\mu\nu}
\eeq

We will now look at three important limits to the correction terms in equation(\ref{brane}) which will help elucidate the range 
of their applicability and their limitations :

\begin{itemize}
 \item {\bf{Low Energy limit}} - As we have already stated the $S_{\mu\nu}$ term makes a significant contribution only at high 
 energy limits $\rho>>\lambda$. However at low energies $\rho<<\lambda$, the correction due to $S_{\mu\nu}$ is negligible 
 and we have only the non-local correction ${\cal E}_{\mu\nu}$.
 \item{\bf{Anti-DeSitter Bulk}} - It can be shown mathematically from the symmetries of the Weyl tensor that it's projection 
 ${\cal E}_{\mu\nu}$=0 if we consider a completely ADS bulk. Then we are left with only the local $S_{\mu\nu}$ correction 
 term.
\item{\bf{Low energy limit, ADS bulk}} - If we take both these limits then both the correction terms become zero and we are left with 
the field equations of General Relativity.
\end{itemize}

\subsubsection{Correction to important physical parameters}

We will now find the forms of the pressure($p$), density($\rho$) and other parameters considering the effective field equations 
(\ref{brane}. Instead of working firsthand with a perfect fluid we first consider a general fluid and then make the perfect 
fluid assumption. This will help us in identifying more clearly the non-local and local correction terms. The E-M tensor for such a fluid 
is :
\beq
\label{emt}
T_{\mu\nu}= \rho u_{\mu} u_{\nu} + p h_{\mu\nu} + \pi_{\mu\nu}+ q_{\mu} u_{\nu} + q_{\nu}u _{\mu}
\eeq
where $u_{\mu}$ is the four velocity, $\pi_{\mu\nu}$ is the anisotropic stress, $q_{\mu}$ the momentum density and 
$h_{\mu\nu}= g_{\mu\nu} + u_{\mu} u_{\nu}$

Using equation (\ref{emt}) we get for the local term $S_{\mu\nu}$ :
\begin{eqnarray}
S_{\mu\nu}= \frac{1}{24}[2\rho^2-3\pi_{\alpha\beta}\pi^{\alpha\beta}]\;u_{|mu}u_{\nu}+\frac{1}{24}[2\rho^2+4\rho p+\pi_{\alpha\beta}
\pi^{\alpha\beta}-4 q_{\alpha}q^{\alpha}]\; h_{\mu\nu} \nonumber \\ -\frac{1}{12}(\rho+3p)\;\pi_{\mu\nu} - \frac{1}{4} \pi_{\alpha}
{{}_{<\mu}\pi_{\nu>}}^{\alpha} - \frac{1}{4} q_{<\mu}q_{\nu>}+   \frac{1}{3}\rho q_{(\mu}u_{\nu)}- \nonumber \\\frac{1}{2} q^{\alpha}
\pi_{\alpha}{}_{(\mu}u_{\nu )}
\end{eqnarray}
where () denotes symmetrization and $<$ $>$ denotes the traceless, symmetric part of the tensors.

For a perfect fluid we have $\pi_{\mu\nu}=0=q_{\mu\nu}$ and $S_{\mu\nu}$ reduces to
\beq
\label{sf}
S_{\mu\nu}=\frac{1}{12} \rho[\rho u_{\mu} u_{\nu} +(\rho+ 2p) h_{\mu\nu}]
\eeq
The form of ${\cal E}_{\mu\nu}$ is :
\beq
\label{ef}
{\cal E}_{\mu\nu}=-\kappa^2[\rho_\varepsilon(u_\mu u_\nu +{1\over 3} h_{\mu\nu}) + q_\mu{}^\varepsilon u_\nu+q_\nu{}^\varepsilon u_\nu
+ \pi_{\mu\nu}^\varepsilon]
\eeq
where $\rho_\varepsilon,p_\varepsilon$ and $\pi_{\mu\nu}{}^\varepsilon$ are the energy density, pressure, momentum density and the anisotropic stress due to non-local effects.

From equations (\ref{sf}, \ref{ef}) the total corrections to the pressure,  density, momentum density and anisotropic stress are 
(for a perfect fluid)
\begin{eqnarray}
\rho_{tot}&=&\rho(1+\frac{\rho}{2\lambda} + \frac{\rho_{\varepsilon}}{\rho}) \\
p_{tot}&=&p+ \frac{\rho}{2\lambda}( 2p+ \rho) +\frac{\rho_{\varepsilon}}{3} \\
q_\mu{}^{tot}&=&q_\mu{}^{\varepsilon} \\
\pi_{\mu\nu}{}^{tot}&=&\pi_{\mu\nu}{}^{\varepsilon}
\end{eqnarray}

The $\rho/\lambda$ terms are the local corrections due to $S_{\mu\nu}$ and the $\rho_\varepsilon$ terms are the non local 
corrections. We can see that even for a perfect fluid we have anisotropic stress (which tells us that isotropy is 
broken) and momentum density (which tells us that spatial homogeneity is broken) terms due to non-local 
effects which is in complete contrast to the analogous expressions in Einstein's G.R. Therefore this can be an important tool 
for detecting Brane World effects as any anisotropic stress term must have an effect on the CMB power spectrum.

The total equation of state is given by :
\beq
\omega_{tot}=\frac{p_{tot}}{\rho_{tot}}= \frac{\omega+(1+2\omega)\frac{\rho}{2\lambda}+\frac{\rho_\varepsilon}{3\rho}}
{1+\frac{\rho}{2\lambda}+\frac{\lambda_\varepsilon}{\rho}}
\eeq
and the sound speed is :
\beq
c_{tot}^2=\frac{\dot p_{tot}}{\dot\rho_{tot}}=\left[c_s^2+\frac{\rho+p}{\rho+2\lambda}+\frac{4\rho_{\varepsilon}}{9(\rho+p)(1+\frac
{\rho}{\lambda}})\right]\left[1+\frac{\rho_{\varepsilon}}{3(\rho+p)(1+\frac{\rho}{\lambda}}\right]^{-1}
\eeq

Here $\omega=\frac{p}{\rho}$ and $c_s^2=\frac{\dot p}{\dot\rho}$ are the ordinary equation of state and sound speed.

At high energies $\rho>>\lambda$ we can ignore the contribution of the non-local effects ($\rho_{\varepsilon}$) and in that limit 
we get
\begin{eqnarray}
\omega_{tot}&\approx &2\omega+1 \\
c_{tot}^2&\approx& c_s^2 +\omega + 1
\end{eqnarray}
 Thus we can see that we have significant corrections to the above parameters in the high energy limit.

\subsubsection{Correction to the Friedmann Equations}

The Friedmann equations can be derived by assuming a FRW metric in the bulk. The exact 
analysis is done in \cite{fr2}, \cite{fr3}. What is important in the context of this report is the correction to the 
Friedmann equations from the standard G.R. case. The modified Friedmann equations are \cite{fr1} :
\beq
\label{friedb}
H^2= \frac{\Lambda}{3} + \left(\frac{8\pi}{3 M_4^2}\right)\rho + \left(\frac{4\pi}{3 M_5^3}\right)^2\rho^2 +\frac{\varepsilon}{a^4}
\eeq

where $M_4^2=\frac{8\pi}{\kappa^2}$ , $M_5^3=\frac{8\pi}{\kappa_5^2}$ and 
$\Lambda=\frac{4\pi}{M_5^3}\left(\Lambda_5+\frac{4\pi}{3 M_5^3}\lambda^2\right)$ when written in terms of the brane tension $\lambda$ 
and the 5-D Planck mass $M_5$

The brane tension $\lambda$ can be constrained by the fact that the the G.R. results work very well from the era of nucleosynthesis 
and so any high energy correction if it's there must come before the energy scale of nucleosynthesis. Thus the term $\frac{\rho}
{\lambda}$ must be insignificant at the energy scale of nucleosynthesis \cite{Maar1}. From this considerations we can get a 
lower bound on the value of $\lambda$ that is
\beq
\label{lambda}
\lambda\geq(1MeV)^4
\eeq
The mathematical value of the Planck Mass in 4-D $M_4$ is :
\beq
M_4=10^{27}eV  
\eeq
and thus the value of the 5-D Planck Mass is $M_5$ :
is 
\beq
M_5=10^{11}eV 
\eeq 
as $M_4$ and $M_5$ are related by :
\beq
M_4=\sqrt{\frac{3}{4\pi}}\left(\frac{M_5^2}{\sqrt{\lambda }}\right)M_5
\eeq
which can be seen from equation(\ref{k2}) and the definitions of $M_4$ and $M_5$.

Now coming back to the correction terms in the equation (\ref{friedb}) we see that the first correction term is $\rho^2$ which 
is due to the local $S_{\mu\nu}$ correction term and the second correction term ($\varepsilon$) is due to the projected Weyl 
tensor term ${\cal E}_{\mu\nu}$. However this second term decays as $a^{-4}$ and thus becomes insignificant after a certain time.

In our work we will assume that the $\varepsilon$ term is zero and choose the form of $\Lambda_5$ , $\Lambda_5\approx
-\frac{4\pi\lambda^2}{3M_5^3}$ such that $\Lambda=0$. Physically thus we fine-tune our parameters to make the
4-D cosmological constant to be zero. Then the modified Friedmann equation reduces to :
\beq
\label{friedb1}
\boxed{H^2=\frac{8\pi}{3M_4^2}\rho\left[1+\frac{\rho}{2\lambda}\right]}
\eeq

\subsubsection{Corrections to Slow Roll Inflation}
\label{csr}

We consider slow roll inflation due to a scalar field on the brane \cite{Maar2}. The Klein-Gordon equation still holds as the conservation 
equation (\ref{conb1}) still holds in Brane-World Gravity under the assumptions we work on (empty bulk) :
\beq
\label{kg}
\ddot\phi+ 3H\dot\phi+\frac{dV(\phi)}{d\phi}=0
\eeq

The condition for inflation can be got from equations (\ref{friedb1}, \ref{kg}) and we get :
\beq
\label{src}
\ddot a>0\Rightarrow p<-\left[\frac{\lambda+2\rho}{\lambda+\rho} \right]\frac{\rho}{3}
\eeq
which is a change from the G.R. condition($p<-\frac{\rho}{3})$.\\

We can get back the G.R. for $\frac{\rho}{\lambda}<<1$ and in the high energy limit $\frac{\rho}{\lambda}>>1$ we get $p<-
\frac{2}{3}\rho$ which is a significant change from the standard results.\\

In G.R. for slow roll to hold we require $\dot\phi^2<<V(\phi)$. However in Brane World that changes to :
\beq
\dot\phi^2-V+\frac{\dot\phi^2+2V}{8\lambda}(5\dot\phi^2-2V)<0
\eeq
which in the high energy limit gives
\beq
\label{src1}
\dot\phi^2<\frac{2}{5} V(\phi)
\eeq
In the slow roll limit the Friedman equations (\ref{friedb1}) and K-G equation (\ref{kg}) reduces to :
\beq
\label{src2}
H^2=\frac{8\pi}{3M_4^2} V\left[1+\frac{V}{2\lambda}\right]
\eeq
and
\beq
\label{src3}
\dot\phi\approx -\frac{V'}{3H}
\eeq
where the $'$ denotes derivative w.r.t. $\phi$.

The slow roll parameters change as :
\beq
\label{epsi}
\boxed{\epsilon_v=\frac{M_4^2}{16\pi}\left({\frac{V'}{V}}\right)^2\left[\frac{1+\frac{V}{\lambda}}{{\left(1+\frac{V}{2\lambda}\right)}^2}\right]}
\eeq
and
\beq
\label{etaa}
\boxed{\eta_v=\frac{M_4^2}{8\pi}\left(\frac{V''}{V}\right)\left[\frac{2\lambda}{2\lambda+V}\right]}
\eeq
The change from standard G.R. based calculations is seen in the high energy as both the parameters are suppressed by a factor by a factor of 
$\frac{V}{\lambda}$. The number of e-folds is given by :
\beq
\boxed{N\approx-\frac{8\pi}{M_4^2}\int_{\phi_i}^{\phi_f} \frac{V}{V'}\left[1+\frac{V}{2\lambda}\right] d\phi}
\eeq
Again the Brane-World effects conspire to increase the number of e-folds due to an extra term of $\frac{V}{\lambda}$. 
This results in more amount of inflation between two field values compared to standard G.R. results.

\subsection{Cosmological Perturbation Theory on the Brane}
\label{cptb}

We follow the same route that we took for cosmological perturbation theory in G.R.(\ref{copeth}). The 
general perturbed metric is the same as that does not change in Brane-World gravity \cite{Maar2}. Thus we can define the same gauge invariant quantities that we defined 
in G.R.. However we have to ensure that the quantities thus defined have zero time evolution on superhorizon scales. This 
is ensured by the conservation equation (\ref{conb1}) which still holds in Brane-World gravity. Here again we will consider that 
${\cal E}_{\mu\nu}=0$ and there is no interaction between the brane and the bulk.

The gauge invariant quantity is defined as :
\beq
\zeta=\Psi-\frac{H}{\dot\rho}\delta\rho
\eeq
Again we have $\dot\zeta\approx 0$ for superhorizon scales and adiabatic perturbations.

In terms of the scalar field and field fluctuations the curvature perturbation on uniform density hypersurfaces is given as :
\beq
\zeta=\frac{H\delta\phi}{\dot\phi}
\eeq
Doimg the same analysis as in G.R. we have for the various cosmological parameters :

The amplitude of the scalar fluctuations :
\beq
\boxed{A_s^2\approx \left(\frac{512\pi}{75M_4^6}\right)\left[\frac{V^3}{V'^2}\left[\frac{2\lambda+V}{2\lambda}\right]^3\right]_ {k=aH}}
\eeq
As we can see, the amplitude is increased at high energies due to the $\frac{V}{\lambda}$ term.

The scale dependence of the perturbation is given as :
\beq
\boxed{n_s-1=\frac{d\ln A_s^2}{d\ln k}\approx -6\epsilon_v+2\eta_v}
\eeq

However as both $\epsilon_v$ and $\eta_v$ are suppressed at high energies we see that in the Brane World context $n_s$ is very close 
to 1 and we get the Harrison-Zel'dovich spectrum ($n_s=1$)

The running of the spectral index ($\alpha_s$) is given as :
\beq
\boxed{\alpha_s= \frac{dn_s}{d\ln k}\approx -24\epsilon_v^2 + 16\epsilon_v\eta_v - 2\xi_v^2}
\eeq

The amplitude of tensor perturbations and the tensor spectral index are :
\beq
A_T^2\approx\frac{32}{75M_4^4}\left[V\left[\frac{2\lambda+V}{2\lambda}\right]\right]_{k=aH}
\eeq
and
\beq
n_T=\frac{d\ln A_T^2}{d\ln k}\approx -2\epsilon_v
\eeq

Again the amplitude of the tensors is increased due to the $\frac{V}{\lambda}$ term while $n_T$ is suppressed as $\epsilon_v$ 
is suppressed. The tensor to scalar ratio is :
\beq
\boxed{r=\frac{A_T^2}{A_s^2}\approx \left[\epsilon_v\left[\frac{\lambda}{\lambda+V}\right]\right]_{k=aH}}
\eeq
Thus we can see that even though both the amplitudes of tensor and scalar perturbations increases the value of $r$ decreases 
drastically due to the $\frac{V}{\lambda}$ term. This is one of the key predictions of Brane-World gravity.\\\\

We see in this chapter that theoretically we get the following predictions in brane-world :
\begin{itemize}
 \item {Value of spectral index pushed towards 1.( signifying Harrison Zel'dovich spectrum)}
 \item{Value of tensor to scalar ratio is found to be very small.}
 \item{Presence of anisotropic stress and momentum density terms even for a perfect fluid.}
 \item{The equation of state parameter and sound speed is changed from the corresponding versions of G.R.} 
\end{itemize}

In the next chapter we will work with certain inflationary models with potential $V(\phi)$ and see if the theoretical predictions hold true 
generally for the various models we will consider. We will also look for discernible differences from standard G.R. results 
which will tell us about the effect of the Brane World corrections and will probably tell us whether there is any need for modification 
in the underlying theory itself.

\chapter{Inflation in Brane-World}
\section{Introduction}

The previous two chapters have delved into the theories of Cosmological perturbation in General Relativity and Brane-World gravity 
. We have given a brief introduction of the above mentioned topics starting with the first chapter where we discussed the standard problems of Big 
Bang Cosmology, the proposed resolution to these difficulties and finally the mathematical details of Cosmological perturbation 
theory and inflation. We have seen what are the observationally relevant parameters and have shown the well founded techniques 
of getting these parameters theoretically. In the next chapter (chapter \ref{braneworld}) Brane World gravity is introduced, the effective 
field equations are derived and the salient, distinctive features of Brane World gravity are elucidated. We then move on to 
Cosmological perturbation theory on the brane and calculate again the observationally relevant parameters. Finally we have the tools at our 
disposal to perform a brane world calculation of Cosmological perturbation given a particular inflationary potential $V(\phi)$.

\section{Modus Operandi}
In this chapter we will work out the prediction of certain inflationary potentials, the forms of which have been well researched 
\cite{martin} in the literature and have been subjected to a rigorous G.R. based analysis. The inflationary potentials chosen are those 
which just exceed the cut-off of the most well favoured inflationary models after Planck(2015) \cite{ring} and the ones which have the 
least number of unconstrained parameters. The reasoning for this is two fold and we will make clear why we make this choices. 
Firstly, we want to see whether the models which are in slight tension with the Planck results can be incorporated into the most 
favourable category due to Brane-World corrections. We do not consider those models which are well favoured by the Planck results. 
Though for the sake of completeness we consider one model of the well-favoured type and see whether whether it is ruled in or ruled 
out. Secondly, we consider minimum no. of unconstrained parameters so that the model doesn't lose it's predictive power. By 
unconstrained parameters we mean parameters which can not be constrained by any other theory or any experimental data and can
thus be changed by hand to fit the experimental results. We also consider some models which are widely ruled out by Planck and see what 
their status is in Brane-World gravity.

The chosen models are :
\begin{itemize}
 \item {\bf{Large Field Inflation- $V(\phi)= M^4(\frac{\phi}{M_4})^p$}} \\\\ These models are discussed from a G.R. based analysis 
 in \cite{martin}. Here $M_4$ is the four dimensional Planck Mass, $p$ can take integer or fractional values. The $M^4$ 
 term is constrained by the scalar perturbation amplitude value which has been precisely measured. A special case of p=2 is the standard 
 chaotic inflation model proposed by Linde \cite{Lind1}. The value of p=2/3 is related to the monodromy potential\cite{mono}. 
 The rest of the values are phenomenological theories and do not have any high energy motivations.
 \item{\bf{Power Law Inflation}}- $V(\phi)= M^4\exp{\left(-\frac{\alpha\phi}{M_4}\right)}$ \\\\ We have $\alpha$ here which is a dimensionless 
 parameter and a positive coefficient. This model was introduced in \cite{pow} \cite{pow2} \cite{pow3} and subjected 
 to a G.R. based analysis in \cite{martin}. It is a phenomenological model.
 \item{\bf{Inverse Monomial Inflation}}- $V(\phi)= M^4\left(\frac{\phi}{M_4}\right)^{-p}$ \\\\ Here again $p$ is a positive number. 
These models are  discussed in \cite{martin} \cite{in1} \cite{in2} \cite{in3}. This is a phenomenological model.
\item{\bf{Open String Tachyonic Inflation}}- $V(\phi)= -M^4\left(\frac{\phi}{\phi_0}\right)^2\ln\left[\left(\frac{\phi}{\phi_0}\right)\right]^2$\\\\
This model is motivated by theoretical considerations and the parameter  $\phi_0= M_s$ where $M_s$ is the string scale which is almost equal to the 4-D Planck scale. 
However if we consider $\phi_0\neq M_s$ then we can remove the high energy underpinnings of the model and it becomes a 
phenomenological model. This model is discussed in \cite{os1} \cite{martin}.

\end{itemize}
 Studying these varying classes of models we will look first at whether they are ruled in or ruled out in Brane World gravity 
 based cosmological perturbation calculations. Secondly we will look at the predictions of theses models and see whether there 
 is an unifying general theme to the predictions for some parameters (independent of models) and understand what the implications are 
 for Brane World gravity. Finally we will present a summary of our results and make our inferences from them.
 
 From our discussions in the previous chapter we have already seen that Brane-World correction give a negligible tensor-to-scalar 
 ratio and a value of the spectral index($n_s$) pushed towards 1( scale invariant spectrum). We will look out for these 
 and see whether these generic predictions stand the test of individual model dependent calculations. Note that a spectral 
 index equal to 1 is strongly ruled out by the recent Planck results \cite{Pl} .
 
 Now we will start our analysis by considering one of the simplest models of chaotic inflation. We will then proceed to the more 
 general potentials.

\section{High Energy Expressions}
We have already seen the forms of the slow roll parameters and the cosmological observables in Brane World gravity in the previous 
sections ({\ref{csr}, \ref{cptb}}). We are interested only in the high energy regime ($\rho>>\lambda$)
here as the Brane World corrections are most 
significant in that regime. We will work exclusively with the Brane World corrections here and ignore the standard G.R. terms in most cases. 
The high energy regime that we consider is suitable for this approximation.
However as it turns out that may not always be feasible and some potentials do need the G.R. terms as we will see later on. Now 
we will list the expressions of the slow roll parameters and the cosmological observables in the high energy regime. We include 
the third slow roll parameter $\xi_v^2$ (\ref{hexi}) here which plays an important role in finding out the running of the spectral 
index (\ref{healph}) which is an important experimental observable and whose determination helps constrain the various inflationary 
models. We list the expressions as follows :\\

{\bf{Slow Roll Parameters}}
\beq
\label{hee}
\epsilon_v\approx \frac{\lambda M_4^2}{4\pi}\frac{V'^2}{V^3}
\eeq
\beq
\label{heeta}
\eta_v\approx \frac{\lambda M_4^2}{4\pi}\left(\frac{V''}{V^2}\right)
\eeq
\beq
\label{hexi}
\xi_v^2\approx \frac{\lambda^2 M_4^2}{16\pi^2}\frac{V'V'''}{V^4}
\eeq

{\bf{Slow Roll Condition}} :
\beq
\label{hesrc}
\epsilon_v\approx 1 \Leftrightarrow \dot\phi^2<\frac{2}{5} V(\phi)
\eeq

{ \bf{No. of e-folds}} :
\beq
\label{hene}
N\approx -\frac{4\pi}{\lambda M_4^2}\int_{\phi_i}^{\phi_f} \frac{V^2}{V'} d\phi
\eeq

{\bf{Cosmological Observables}}\\

{\bf{Scalar Amplitude}} :
\beq
\label{heas}
A_s^2\approx\frac{64}{75}\frac{\pi}{\lambda^3 M_4^6}\left(\frac{V'^6}{V^2}\right)_{k=aH}
\eeq

{\bf{Scalar spectral index}} :
\beq
\label{hens}
n_s\approx [1+2\eta_v-6\epsilon_v]_{k=aH}
\eeq

{\bf{Running of the spectral index}} :
\beq
\label{healph}
\alpha_s=[\frac{dn_s}{d\ln k}\approx -24\epsilon_v^2 + 16\epsilon_v\eta_v - 2\xi_v^2]_{k=aH}
\eeq

{\bf{Tensor-to-scale ratio}} :
\beq
\label{herat}
r\approx \left[\frac{\epsilon_v \lambda}{V}\right]_{k=aH}
\eeq

{\bf{Values of}} $M_4,M_5,\lambda$
\begin{eqnarray}
\label{values}
M_4= 10^{28} eV \\ M_5=10^{11}eV \\ \lambda= 10^{24}eV
\end{eqnarray}

Now we will start with the first simple model of chaotic inflation with potential $V(\phi)= m^2\phi^2$. We will later see that 
this is a special case of a more general class of models called Large Field Inflation(LFI)

\section{Inflationary Models}
\subsection{A First Model: $V(\phi)= m^2\phi^2$}

\begin{figure}[h!!!!]
\centering
\epsfig{figure=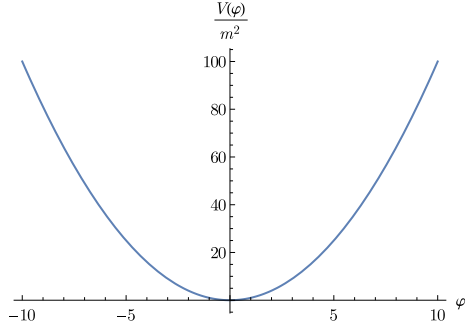,width=6cm,angle=0}
\caption{{\bf Chaotic Potential} : Slow roll proceeds for positive field values in the direction of decreasing $\phi$. The 
parameter $m$ is found out to be $\approx 10^{-2}M_5$.}
\label{chaotic}
\end{figure}

Here $m$ is just a parameter constrained by the scalar perturbation and the potential is shown in Fig: (\ref{chaotic})

We first find the derivatives of this potential :
\begin{eqnarray}
\label{dech}
V'&=&2m^2\phi \nonumber \\ V''&=&2m^2\nonumber \\ V'''&=&0
\end{eqnarray}

The {\bf{Slow Roll Parameters}} are :
\begin{eqnarray}
\label{ce}
\epsilon_v\approx\frac{\lambda M_4^2}{4\pi}\frac{V'^2}{V^3} \approx\frac{\lambda M_4^2}{\pi}\frac{1}{m^2\phi^4}
\end{eqnarray}

\begin{eqnarray}
\label{ceta}
\eta_v\approx\frac{\lambda M_4^2}{4\pi}\left(\frac{V''}{V^2}\right)
\approx\frac{\lambda M_4^2}{2\pi}\frac{1}{m^2\phi^4} \approx\frac{\epsilon_v}{2}
\end{eqnarray}

and

\begin{eqnarray}
\label{cxi}
\xi_v^2\approx\frac{\lambda^2M_4^4}{16\pi}\frac{V' V'''}{V^4}=0 
\end{eqnarray}

{\bf{No. of e-folds}}
\begin{eqnarray}
\label{cno}
N\approx-\frac{4\pi}{\lambda M_4^2} \int_{\phi_i}^{\phi_{end}} \frac{V^2}{V'} d\phi \approx-\frac{2\pi m^2}{\lambda M_4^2}\left[\frac{\phi^4}{4}\right]_{\phi_{i}}^{\phi_{end}} \approx\frac{\pi m^2}{2\lambda M_4^2}\left[{\phi_{i}}^4-{\phi_{end}}^4\right]
\end{eqnarray}

The condition for slow roll in the high energy limit is given as equation (\ref{hesrc}). It can be seen by explicit 
calculations that for both these conditions we get the same expression except a factor of 1.2 and so we will work with the 
condition $\epsilon_v\approx 1$. Thus to get $\phi_{end}$ we require violation of the slow roll condition :
\beq
\label{pend}
\frac{1}{m^2\phi_{end}^4}\approx\frac{\pi}{\lambda M_4^2}
\eeq

Now putting this value of $\phi_{end}$ from equation (\ref{pend}) in equation (\ref{cno}) we get :
\beq
N\approx \frac{\pi m^2}{2\lambda M_4^2} \phi^4_{i} - \frac{\pi m^2}{2\lambda M_4^2} \phi_{end}^4
\eeq

The value of the no. of e-folds $N$ is taken as 58 and therefore we have :
\begin{eqnarray}
\label{c2}
\frac{\pi m^2}{2\lambda M_4^2} \phi_{*}^4&\approx& 58+ \frac{\pi m^2}{2\lambda M_4^2}\phi_{end}^4 \approx
58+ \frac{1}{2}  \nonumber \\ m^2&\approx& \left(\frac{2\lambda M_4^2}{\pi}\right)\times
\frac{58.5}{\phi_{*}^4}
\end{eqnarray}
where $\phi_{*}$ is the value of the field variable at horizon crossing $k=aH$.

To constrain the values of $m$ in our model we turn to the scalar power spectrum $A_s^2$ which has been measured precisely 
and we use it to fix the value of $m$  to get the appropriate scalar power spectrum which is from equations(\ref{heas}, \ref{dech}) :
\begin{eqnarray}
\label{c3}
 A_s^2\approx\frac{64}{75} \frac{\pi}{M_4^6 \lambda^3}\left(\frac{m^{12} \phi_{*}^{12}}{4 m^4 \phi_{*}^2}\right) \approx\frac{16\pi}{75}\frac{1}{\lambda^3 M_4^6}\times m^8 \phi_{*}^{10}
\end{eqnarray}

Now putting the value of $m^2$ (\ref{c2}) in equation (\ref{c3}) we get :
\begin{eqnarray}
A_S^2&\approx& 1.28\times 10^6\times \lambda M_4^2\times \frac{1}{\phi_{*}^6}
\end{eqnarray}
From Planck data(\cite{Pl}) we get we get :
\beq
A_s^2= 2.23\pm0.16\times 10^{-9}
\eeq

Thus-
\begin{eqnarray}
 \phi_{*}^6&\approx&\frac{1.28\times10^6\times \lambda M_4^2}{2.23\times 10^{-9}} \nonumber \\ 
 \phi_{*}&\approx& 6.2108\times 10^{4} M_5
\end{eqnarray}

and thus from equation(\ref{c2}) we get :
\beq
m^2\approx 2.4993\times 10^{7}M_5
\eeq

Now with these values we can get the spectral index(\ref{hens}) and the running of the spectral index(\ref{healph}) :
\begin{eqnarray}
\label{c5}
n_s\approx \left[1+2\eta_v-6\epsilon_v\right]_{k=aH} 
\end{eqnarray}
\begin{eqnarray}
 \alpha_s=\left[\frac{dn_s}{d\ln k}\approx -24\epsilon_v^2 + 16\epsilon_v\eta_v 
- 2\xi_v^2\right]_{k=aH}
\label{c7}
\end{eqnarray}

where
\begin{eqnarray}
 \label{c4}
\epsilon_v&\approx&\frac{\lambda M_4^2}{\pi}\frac{1}{m^2\phi_{*}^4} \approx\frac{\lambda M_4^2}{\pi}
\times\frac{{\pi}}{{2\lambda M_4^2\times 58.5}} \nonumber \\&=&8.5592\times 10^{-3}
\end{eqnarray}
and
\beq
\label{c6}
\eta_v=\frac{\epsilon_v}{2}=4.2796\times 10^{-3}
\eeq
Now putting these(\ref{c4},\ref{c6}) values in equations(\ref{c5}) we get :
\begin{eqnarray}
\boxed{n_s=1-5\epsilon_v =0.9571 }
\end{eqnarray}
and equation(\ref{c7})
\begin{eqnarray}
\boxed{\alpha_s=-24\epsilon_v^2+16\times\epsilon_v^2-0 =-16\epsilon_v^2=1.1722\times 10^{-3}}
\end{eqnarray}

The tensor to scalar ratio is found to be :
\begin{eqnarray}
 \label{c8}
 \boxed{r=\left[\frac{\epsilon_v\times \lambda}{V}\right]_{k=aH} \approx 10^{-26}}
\end{eqnarray}

as we can see it is a fantastically small number.

Now that we have finished working with a comparatively simpler model we start now with a model of which the previous worked out 
model was just was a special case. This model with $p$=1,2,2/3 just misses the cutoff of Planck 2015 \cite{Pl}. We will calculate the 
predictions of this model in Brane-World gravity for these values of $p$ and some other values which are phenomenologically viable.

\subsection{Large Field Inflation : $V(\phi)=M^4\left(\frac{\phi}{M_4}\right)^p$}

\begin{figure}[h!!!!]
\centering
\epsfig{figure=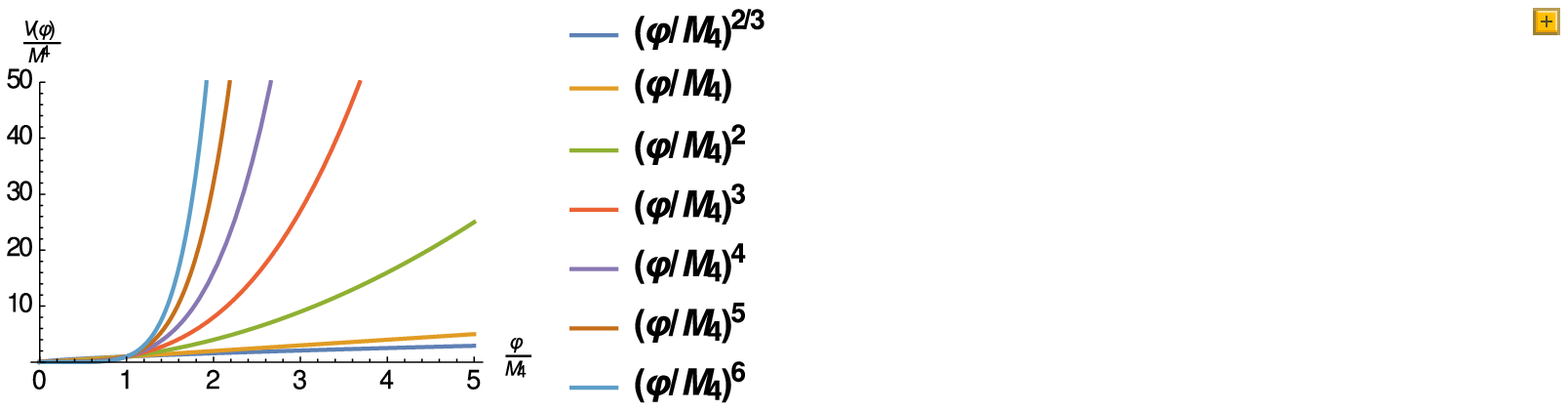,width=10cm,height=7cm,angle=0}
\caption{{\bf Large Field Inflation} : The slope of the potential becomes steeper as the power increases and 
slow roll takes place in the direction of decreasing field values. The value of $M^4$ is given by Equation (\ref{M4lf}).}
\label{lfip1}
\end{figure}

The derivatives of the potential are :
\begin{eqnarray}
\label{delfi}
V'&=&\frac{M^4}{M_4}p\left(\frac{\phi}{M_4}\right)^{p-1}\nonumber \\ V''&=&\frac{M^4}{M_4^2}p(p-1)\left(\frac{\phi}{M_4}\right)^{p-2}
\nonumber \\ V'''&=&\frac{M^4}{M_4^3}p(p-1)(p-2)\left(\frac{\phi}{M_4}\right)^{p-3}
\end{eqnarray}

The {\bf{Slow Roll Parameters}} are from equations (\ref{hee}, \ref{heeta}, \ref{hexi}, \ref{delfi}) :
\begin{eqnarray}
\label{le}
\epsilon_v\approx \frac{\lambda M_4^2}{4\pi}\frac{V'^2}{V^3}  \approx
\frac{\lambda p^2}{4\pi M^4}\left(\frac{\phi}{M_4}\right)^{-(p+2)}
\end{eqnarray}
and
\begin{eqnarray}
\label{leta}
\eta_v\approx \frac{\lambda M_4^2}{4\pi}\left(\frac{V''}{V^2}\right) 
\approx \frac{\lambda p(p-1)}{4\pi M^4}\left(\frac{\phi}{M_4}\right)^{-(p+2)}
\end{eqnarray}
and
\begin{eqnarray}
\label{lxi}
\xi_v^2\approx \frac{\lambda^2 M_4^2}{16\pi^2}\frac{V'V'''}{V^4} \approx\frac{\lambda^2 p^2 (p-1)(p-2)}{16\pi^2 M^8}\left(\frac{\phi}{M_4}\right)^{-2(p+2)}
\end{eqnarray}

The {\bf{No. of e-folds}} :
\begin{eqnarray}
N&\approx& -\frac{4\pi}{\lambda M_4^2}\int_{\phi_{i}}^{\phi_{end}} \frac{V^2}{V'} d\phi \approx
-\frac{4\pi M^4}{\lambda p(p+2)}\left[\left(\frac{\phi}{M_4}\right)^{p+2}\right]
_{\phi_{i}}^{\phi_{end}} \nonumber \\ &\approx&\frac{4\pi M^4}{\lambda p(p+2)}\left[\left(\frac{\phi_{i}}{M_4}\right)^{p+2} -
\left(\frac{\phi_{end}}{M_4}\right)^{p+2}\right]
\label{l1}
\end{eqnarray}

Again the condition for end of slow roll is $\epsilon_v\approx 1$ . Therefore we have from equation (\ref{le}) :

\beq
\label{l2}
\left(\frac{\phi_{end}}{M_4}\right)^{p+2}\approx \frac{\lambda p^2}{4\pi M^4}
\eeq

Now putting this value of equation ({\ref{l2}) in equation ({\ref{l1})
\begin{eqnarray}
N&\approx& - \frac{p}{p+2} +\frac{4\pi M^4}{\lambda p(p+2)}\left(\frac{
\phi_{i}}{M_4}\right)^{p+2} \nonumber \\ M^4 &\approx& \frac{\lambda}{\pi}\left[58 p(p+2) +5 p^2\right]\left(\frac{\phi_{*}}
{M_4}\right)^{-(p+2)}
\label{l3}
\end{eqnarray}\\
where $N=58$ and $\phi_{*}$ is the value of $\phi$ at horizon crossing $k=aH$, i.e. ${[(\phi)_{k=aH}=\phi_{*}]}$\\

To constrain the value of $M$ we turn again to the scalar power spectrum.
\begin{eqnarray}
A_s^2 \approx \frac{64}{75} \frac{\pi}{\lambda^3 M_4^6}\left(\frac{V^6}{V'^2}\right)_{k=aH} \approx \frac{64}{75} \frac{\pi}{\lambda^3 M_4^4 p^2} M^{16} \left(\frac{\phi_{*}}{M_4}\right)^{4p+2}
\label{l4}
\end{eqnarray}

Putting the value of $M^4$ (\ref{l3}) in the equation(\ref{l4}) :
\begin{eqnarray}
A_s^2&\approx&\frac{64}{75} \frac{\pi}{\lambda^3 M_4^4 p^2}\times (\frac{\lambda}{\pi})^4[58p(p+2) +  p^2]^4
\left(\frac{\phi_{*}}{M_4}\right)^{-(4p+8)}\left(\frac{\phi_{*}}{M_4}\right)^{4p+2} \nonumber \\ A_s^2&\approx&\frac{
2.75\times 10^{-2}}{M_4^4}\frac{\lambda}{p^2}[58p(p+2)+p^2]^4\left(\frac{\phi_{*}}{M_4}\right)^{-6} \nonumber \\
\frac{\phi_{*}}{M_4}&\approx&\left[\frac{1.23\times 10^6[58p(p+2)+ p^2]^4}{p^2}\right]^{1/6} \times 4.64\times 10^{-14}
\label{l5}
\end{eqnarray}
and thus we get $M^4$ from equations (\ref{l3},\ref{l5}) :
\beq
\label{M4lf}
M^4 \approx \frac{\lambda}{\pi}\left[58 p(p+2) +5 p^2\right]\left[\left(\frac{1.23\times 10^6[58p(p+2)+ p^2]^4}{p^2}\right)^{1/6} 
\times 4.64\times 10^{-14}\right]^{-(p+2)}
\eeq
 Now we look at the predictions of this models for the spectral index and the running of the spectral index. The forms of the 
 slow roll parameters are from equations (\ref{le}, \ref{leta}, \ref{lxi}, \ref{l3}) :
 \begin{eqnarray}
 \epsilon_v=\frac{\lambda p^2}{4\pi}\left[\frac{24\pi}{\lambda[58p(p+2)+ p^2]}\right] = \frac{p^2}{[58p(p+2)+p^2]} 
 \end{eqnarray}
and
\begin{eqnarray}
\eta_v=\frac{\lambda p(p-1)}{4\pi}\left[\frac{\pi}{\lambda[58p(p+2)+p^2]}\right] =\frac{p(p-1)}{[58p(p+2)+p^2]} 
\end{eqnarray}
and
\begin{eqnarray}
\xi_v^2=\frac{\lambda^2 p^2(p-1)(p-2)}{16\pi^2}\left[\frac{\pi}{\lambda[58p(p+2)+p^2]}\right]^2 = \frac{
p^2(p-1)(p-2)\times 6.25\times 10^{-2}}{[58p(p+2)+p^2]^2}
\end{eqnarray}
 Thus the spectral index is :
 \begin{eqnarray}
 \label{lfi1e}
n_s&\approx& 1+2\eta_v-6\epsilon_v \approx \boxed{1+\frac{2(1-\frac{1}{p})-6}{[58(1+\frac{2}{p})+1]}}
 \end{eqnarray}
 The running of the spectral index is :
 \begin{eqnarray}
 \label{lfi1a}
\alpha_s\approx-24\epsilon_v^2 + 16\epsilon_v\eta_v - 2\xi_v^2 \approx \boxed{\frac{-24+16(1-\frac{1}{p})- 125(1-\frac{1}{p})
(1-\frac{2}{p})}{[58(1+\frac{2}{p})+1]^2}}
 \end{eqnarray}
 
 The tensor to scalar ratio comes out to be :
 \begin{eqnarray}
 \label{lfi1r}
r&\approx&\left[\epsilon_v\times\frac{\lambda}{V}\right]_{k=aH} \approx \boxed{\frac{1.03\times 10^{-24}}{[58(1+\frac{2}{p})+1]^{2/3}}}
 \end{eqnarray}

 Now we look at specific models of this potential :

\subsubsection{$p$=2/3 : $V(\phi)=M^4\left(\frac{\phi}{M_4}\right)^{2/3}$}

\begin{itemize}
 \item {\bf{Spectral Index}}: ($n_s)=0.9699$
 \item{\bf{Running of the spectral index}}: ($\alpha_s)= -5.9028\times 10^{-4}$
 \item{\bf{Tensor-to-scalar ratio}}: ($r)=8.24\times 10^{-27}$
\end{itemize}
\subsubsection{$p$=1 : $ V(\phi)=M^4\left(\frac{\phi}{M_4}\right)$}

\begin{itemize}
 \item {\bf{Spectral Index}}: ($n_s)=0.9672$
 \item{\bf{Running of the spectral index}}: ($\alpha_s)= -7.8361\times 10^{-4}$
 \item{\bf{Tensor-to-scalar ratio}}: ($r)=9.97\times 10^{-27}$
\end{itemize}

\subsubsection{$p=2$ : $ V(\phi)=M^4\left(\frac{\phi}{M_4}\right)^{2}$}

\begin{itemize}
 \item {\bf{Spectral Index}}: ($n_s)=0.9572$
 \item{\bf{Running of the spectral index}}: ($\alpha_s)= -1.1688\times 10^{-3}$
 \item{\bf{Tensor-to-scalar ratio}}: ($r)=4.31\times 10^{-26}$
\end{itemize}

\subsubsection{$p=3$ : $ V(\phi)=M^4\left(\frac{\phi}{M_4}\right)^{3}$}

\begin{itemize}
 \item {\bf{Spectral Index}}: ($n_s)=0.9522$
 \item{\bf{Running of the spectral index}}: ($\alpha_s)= -4.3098\times 10^{-3}$
 \item{\bf{Tensor-to-scalar ratio}}: ($r)=4.86\times 10^{-26}$
\end{itemize}

\subsubsection{$p=4$ : $ V(\phi)=M^4\left(\frac{\phi}{M_4}\right)^{4}$}

\begin{itemize}
 \item {\bf{Spectral Index}}: ($n_s)=0.9489$
 \item{\bf{Running of the spectral index}}: ($\alpha_s)= -7.6030\times 10^{-3}$
 \item{\bf{Tensor-to-scalar ratio}}: ($r)=5.2\times 10^{-26}$
\end{itemize}

\subsubsection{$p$=5 : $ V(\phi)=M^4\left(\frac{\phi}{M_4}\right)^{5}$}

\begin{itemize}
 \item {\bf{Spectral Index}}: ($n_s)=0.9465$
 \item{\bf{Running of the spectral index}}: ($\alpha_s)= -10.5371\times 10^{-3}$
 \item{\bf{Tensor-to-scalar ratio}}: ($r)=5.51\times 10^{-26}$
\end{itemize}

\subsubsection{$p$=6: $ V(\phi)=M^4\left(\frac{\phi}{M_4}\right)^{6}$}

\begin{itemize}
 \item {\bf{Spectral Index}}: ($n_s)=0.9447$
 \item{\bf{Running of the spectral index}}: ($\alpha_s)= -13.0561\times 10^{-3}$
 \item{\bf{Tensor-to-scalar ratio}}: ($r)=5.62\times 10^{-26}$
\end{itemize}

\subsection{Power Law Inflation: $V(\phi)=M^4\exp{\left(-\frac{\alpha\phi}{M_4}\right)}$}

\begin{figure}[h!!!!]
\centering
\epsfig{figure=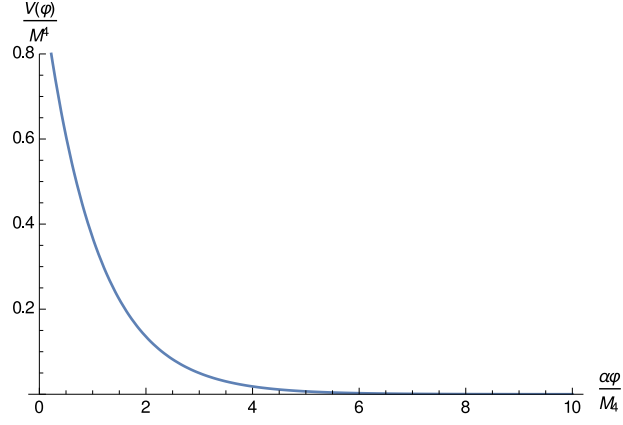,width=8cm,angle=0}
\caption{{\bf Power Law Inflation} : Slow roll inflation proceeds from left to right with increasing field values. The values 
of $\alpha$ and $M^4$ are given by equations (\ref{p4}, \ref{al}).}
\label{plia}
\end{figure}

The derivatives of the potential :
\begin{eqnarray}
\label{depli}
V'=-\frac{\alpha}{M_4} M^4 \exp{\left(-\frac{\alpha\phi}{M_4}\right)}\nonumber \\ 
V''=\frac{\alpha^2}{M_4^2}M^4\exp{\left(-\frac{\alpha\phi}{M_4}\right)}\nonumber \\
V'''=-\frac{\alpha^3}{M_4^3} M^4\exp{\left(-\frac{\alpha\phi}{M_4}\right)}
\end{eqnarray}
The slow roll parameters are (\ref{hee}, \ref{heeta}, \ref{hexi}, \ref{depli}):
\begin{eqnarray}
\label{plie}
\epsilon_v\approx\frac{\lambda M_4^2}{4\pi}\frac{V'^2}{V^3} \approx \frac{\lambda\alpha^2}{4\pi}
\frac{\exp{\left(\frac{\alpha\phi}{M_4}\right)}}{M^4}
\end{eqnarray}
and
\begin{eqnarray}
\eta_v\approx \frac{\lambda M_4^2}{4\pi}\left(\frac{V''}{V^2}\right)  \approx \frac{\lambda
\alpha^2}{4\pi}\frac{\exp{\left(\frac{\alpha\phi}{M_4}\right)}}{M^4} =\epsilon_v
\end{eqnarray}
and
\begin{eqnarray}
\xi_v^2\approx \frac{\lambda^2 M_4^2}{16\pi^2}\frac{V'V'''}{V^4}  \approx \left(\frac{\lambda\alpha^2}{4\pi}\frac{\exp{\left(\frac{\alpha\phi}{M_4}\right)}}{M^4}\right)^2
=\epsilon_v^2
\end{eqnarray}

Therefore we have :
\beq
\label{p1}
\epsilon_v=\eta_v=\sqrt{\xi_v^2}
\eeq

The No. of e-folds is given as :
\begin{eqnarray}
\label{p3}
N&\approx& -\frac{4\pi}{\lambda M_4^2}\int_{\phi_{i}}^{\phi_{end}} \frac{V^2}{V'} d\phi \approx \frac{4\pi M^4}{\lambda M_4\alpha}\int_{\phi_{i}}^{\phi_{end}} \exp{\left(-\frac{\alpha\phi}{M_4}\right)} d\phi
\nonumber \\ &\approx& \frac{4\pi M^4}{\lambda\alpha^2} \left[ \exp{\left(-\frac{\alpha\phi_{i}}{M_4}\right)} - 
\exp{\left(-\frac{\alpha\phi_{end}}{M_4}\right)}\right]
\end{eqnarray}

The condition for inflation to end is $\epsilon_v\approx 1$ and from equation (\ref{plie}) :
\beq
\label{p2}
M^4\exp{\left(-\frac{\alpha\phi_{end}}{M_4}\right)}\approx \frac{\lambda\alpha^2}{4\pi}
\eeq

Now putting value of equation(\ref{p2}) in equation(\ref{p3}) we get on putting $N=58$ :
\begin{eqnarray}
N &\approx&
\frac{4\pi}{\lambda\alpha^2} M^4 \exp{\left(-\frac{\alpha\phi_{i}}{M_4}\right)} - 1  \nonumber \\ 
M^4\exp{\left(-\frac{\alpha\phi_{*}}{M_4}\right)}&\approx& 14.75\times \frac{\lambda\alpha^2}{\pi}
\label{p4}
\end{eqnarray}
where $\phi_{*}$ is the value of the field variable at horizon crossing $k=aH$.

Now we turn to the scalar power spectrum to see whether we can constrain the values of $\alpha$ or $M$ of the potential. That is 
we want to specify the value of $\alpha$ or $M$ which will give us the required amplitude of the spectrum of scalar perturbation 
from equation(\ref{p4})
\begin{eqnarray}
A_s^2&\approx&\frac{64}{75}\frac{\pi}{\lambda^3 M_4^6}\left(\frac{V^6}{V'^2}\right)_{k=aH}  \approx
\frac{64}{75}\frac{\pi}{\lambda^3 M_4^4\alpha^2}\left(M^4\exp{-\left(\frac{\alpha\phi_{*}}{M_4}\right)}\right)^4 
\nonumber \\ &\approx& 1.2280\times10^3\times \frac{\lambda\alpha^6}{M_4^4}\nonumber \\ \alpha^6&\approx& \frac{2.23\times
10^{-9}\times M_4^4}{1.2280\times 10^3 \lambda}\nonumber \\
\alpha&\approx& \boxed{1.1045\times10^{-6}\times (M_4)^{2/3}}
\label{al}
\end{eqnarray}
An interesting feature of this potential is that the term $M$ is not constrained directly by the amplitude of the scalar power 
spectrum. Rather the term $\alpha$ is constrained. However this is not problematic mathematically as from equation(\ref{p4}) 
we can get a form of $M^4$ and $\phi_{*}$ of the form given in equation (\ref{p4}) which is all that we require to complete 
our calculations and specify the particular potential. Now we look at the predictions of this model for the spectral index and 
the running of the spectral index.
\beq
n_s=[1+2\eta_v-6\epsilon_v]_{k=aH}
\eeq
However $\epsilon_v=\eta_v$ we get :
\beq
n_s= 1-4\epsilon_v
\label{lfie1}
\eeq
and
the slow roll parameters are from equations (\ref{p4},\ref{p1},\ref{plie}) :
\begin{eqnarray}
\label{lfie2}
\epsilon_v= \frac{\lambda\alpha^2}{4\pi}\frac{\exp{-\left(\frac{\alpha\phi_{*}}{M_4}\right)}}{M^4} \epsilon_v=0.01695
\end{eqnarray}
Putting value in equation (\ref{lfie1}) we get :
\beq
n_s\approx1-4\times 0.01695\approx\boxed{0.9322}
\eeq
The running of the spectral index is given as considering equation(\ref{p1},\ref{lfie2}) :
\begin{eqnarray}
\alpha_s=\frac{dn_s}{d\ln k}\approx -24\epsilon_v^2 + 16\epsilon_v\eta_v - 2\xi_v^2 \approx -10\epsilon_v^2 \approx \boxed{-2.8866\times 10^{-3}}
\end{eqnarray}
The tensor to scalar ratio is given as considering equation(\ref{al}) :
\begin{eqnarray}
r\approx\left[\epsilon_v\frac{\lambda}{V}\right]_{k=aH} \approx 0.01695\times \frac{10^{24}}{M^4
\exp{-\left(\frac{\alpha\phi_{cmb}}{M_4}\right)}} 
\approx \frac{3.6289\times 10^{-3}}{\alpha^2} \approx \boxed{1.3807\times 10^{-28}}
\end{eqnarray}

An interesting point to be noted here is that the value of $\alpha$ is required only for finding the value of $r$ while the other values of $n_s$ \
and $\alpha_s$ are found without any need to constrain the value of $\alpha$.

\subsection{Inverse Monomial Inflation : $V(\phi)= M^4\left(\frac{\phi}{M_4}\right)^{-p}$}

Here p is any parameter and $M^4$ needs to be constrained by the amplitude if the scalar power spectrum. This model has been ruled 
out comprehensively by a G.R. based analysis. However it is interesting to note what happens when Brane effects are taken into account.

\begin{figure}[h!!!!]
\centering
\epsfig{figure=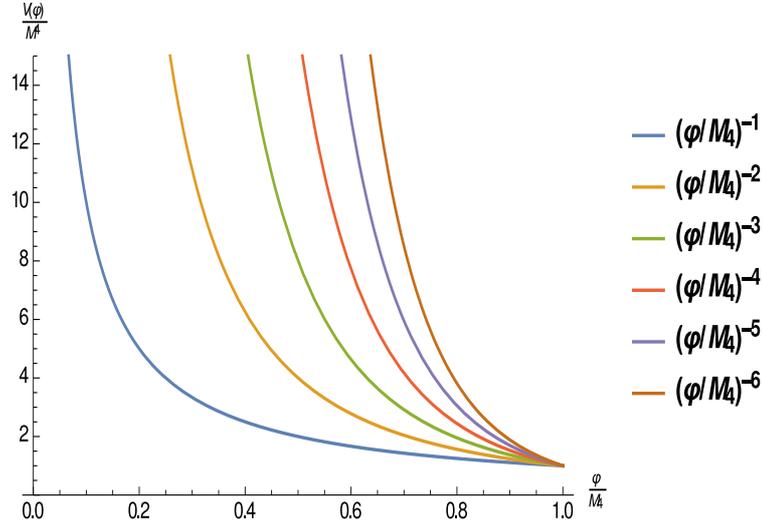,width=10cm,height=7cm,angle=0}
\caption{{\bf Inverse Monomial Inflation} : The potential becomes steeper as the power increases and slow
roll inflation proceeds from low field 
values to high field values (from left to right along the x-axis). The value of $M^4$ is given in equation (\ref{mimi})
}
\label{imip}
\end{figure}

The derivatives of the potential are :
\begin{eqnarray}
\label{deimi}
V'&=&\frac{M^4}{M_4}(-p) \ppl^{-(p+1)} \nonumber \\ V''&=& p(p+1)\frac{M^4}{M_4^2}\ppl^{-(p+2)} \nonumber \\
V'''&=& -p(p+1)(p+2)\frac{M^4}{M_4^3}\ppl^{-(p+3)}
\end{eqnarray}

The slow roll parameters are from equations (\ref{hee}, \ref{heeta}, \ref{hexi}, \ref{deimi}) :
\begin{eqnarray}
\epsilon_v\approx \frac{\lambda M_4^2}{4\pi}\frac{V'^2}{V^3} \approx \frac{\lambda M_4^2}{4\pi}\ppl^{(p-2)}
\label{imie}
\end{eqnarray}
and
\begin{eqnarray}
\label{imieta}
\eta_v\approx \frac{\lambda M_4^2}{4\pi}\left(\frac{V''}{V^2}\right) \approx 
\frac{p(p+1)}{4\pi M^4}\times \ppl^{(p-2)}
\end{eqnarray}
and
\begin{eqnarray}
\label{imixi}
\xi_v^2\approx \frac{\lambda^2 M_4^2}{16\pi^2}\frac{V'V'''}{V^4} 
\approx \frac{\lambda ^2 p^2(p+1)(p+2)}{16\pi^2 M^8}\times \ppl^{(2p-2)}
\end{eqnarray}
The No. of e-folds :
\begin{eqnarray}
N&\approx& -\frac{4\pi}{\lambda M_4{}^2}\int_{\phi_{i}}^{\phi_{end}} \frac{V^2}{V'} d\phi \approx \frac{4\pi M^4}
{\lambda M_4 p}\int_{\phi_{i}}^{\phi_{end}} \ppl^{(-p+1)} d\phi \nonumber \\ &\approx& \frac{4\pi M^4}
{\lambda M_4 p(2-p)} \left[\ppl^{(-p+2)}\right]_{\phi_{i}}^{\phi_{end}} \nonumber \\ &\approx& \frac{4\pi M^4}
{\lambda M_4 p}\left[\pplc^{-(p-2)} - \pple^{-(p-2)} \right]
\label{imin}
\end{eqnarray}
 The condition of end of inflation is $\epsilon_v\approx 1$ and therefore :
 \beq
\label{imi2} 
\pple^{-(p-2)}\approx \frac{4\pi M^4}{\lambda p^2}
\eeq
Putting value of equation(\ref{imi2}) in equation (\ref{imin}) :
\begin{eqnarray}
N &\approx&\frac{4\pi M^4}{\lambda p(p-2)} \pplc^{-(p-2)}-\frac{p}{p-2}  \nonumber \\ \pplc^{-(p-2)} &\approx&
\left(N+\frac{p}{p-2}\right)\frac{\lambda p(p-2)}{4\pi M^4}
\label{imin1}
\end{eqnarray}
To constrain the value of $M$ we turn to the scalar power spectrum :
\begin{eqnarray}
A_s^2&\approx&\frac{64}{75}\frac{\pi}{\lambda^3 M_4^6}\left(\frac{V^6}{V'^2}\right) \approx
\frac{64}{75}\frac{\pi}{\lambda^3 M_4^4p^2} M^{16}\ppls^{(-4p+2)} \nonumber \\ &\approx&
\frac{64}{75}\frac{\pi}{\lambda^3 M_4^4p^2} \times[N+\frac{p}{p-2}]\frac{\lambda^4 p^4(p-2)^4}{(4\pi)^4}\times\ppls^{(-6)} \nonumber \\ \ppls^{6}&\approx& 4.8208\times 10^4 \frac{\lambda p^2(p-2)^4}{M_4^4}\left(58+\frac{p}{p-2}\right)^4
\nonumber \\ \ppls&\approx& 6.0328\left[\frac{\lambda p^2(p-2)^4}{M_4^4}\right]^{1/6} \left(58+\frac{p}{p-2}\right)^{2/3}
\end{eqnarray}
where $\phi_{*}$ is the value of the field variable at horizon crossing $k=aH$.

 We can see that it has different values for different values of $p$.
\beq
\label{mimi}
M^4\approx \left(N+\frac{p}{p-2}\right)\frac{\lambda p(p-2)}{4\pi}
\left[6.0328\left(\frac{\lambda p^2(p-2)^4}{M_4^4}\right)^{1/6} \left(58+\frac{p}{p-2}\right)^{2/3}\right]^{(p-2)}
\eeq 
 We now move on to the predictions of spectral index and running of the spectral index. We evaluate the values of the slow roll parameter 
 at horizon crossing.
 \beq
 \label{lns}
 n_s\approx 1+2\eta_v-6\epsilon_v
\eeq

and
\beq
\label{las}
\alpha_s=\frac{dn_s}{d\ln k}\approx -24\epsilon_v^2 + 16\epsilon_v\eta_v - 2\xi_v^2
\eeq

The slow roll parameters at horizon crossing are found from equations (\ref{imie}, \ref{imieta}, \ref{imixi}, \ref{imin1}) :
\begin{eqnarray}
\label{ine}
\epsilon_v\approx \frac{\lambda p^2}{4\pi M^4}\ppls^{(p-2)} = \frac{p}{(p-2)}\frac{1}{\left[58+\frac{p}{p-2}\right]} 
\end{eqnarray}
and
\begin{eqnarray}
\label{ineta}
\eta_v\approx \frac{\lambda p(p-1)}{4\pi M^4} \ppls^{(p-2)} = \frac{(p+1)}{(p-2)\left(58+\frac{p}{p-2}\right)}
\end{eqnarray}
\begin{eqnarray}
\xi_v&\approx& \frac{\lambda ^2 p^2(p+1)(p+2)}{16\pi^2 M^8}\times \ppls^{(2p-2)} =
\frac{(p+1)(p+2)}{(p-2)^2\left[58+\frac{p}{p-2}\right]^2}
\end{eqnarray}
Therefore we have from equations (\ref{lns}, \ref{las})-
\begin{eqnarray}
n_s\approx \boxed{1+\frac{2-4p}{(p-2)\left(58+\frac{p}{p-2}\right)}} 
\end{eqnarray}
and
\begin{eqnarray}
\alpha_s\approx\boxed{\frac{-24p^2 + 16p(p+1)- 2(p+1)(p+2)}{(p-2)^2\left(58+\frac{p}{p-2}\right)^2}}       
\end{eqnarray}
and
\begin{eqnarray}
r\approx \epsilon_v\frac{\lambda}{V} \approx \boxed{4.5735\frac{\left[\frac{\lambda p^2(p-2)^4}{M_4^4}\right]^{1/3}\times \left(58+\frac{5}{6} 
\frac{p}{p-2}\right)^{1/3}}{p(p-2)}}
\end{eqnarray}

However we now make an interesting comment on the values of $\epsilon_v, \eta_v$ and thus $r$(\ref{herat})
for $p=1,2$. As we can see from the  equations(\ref{ine}, \ref{ineta}) we get negative values for 
$\epsilon_v,r$ for $p=1$ and the equations blow up for $p=2$. These are unphysical results which leads us to the fact that 
the high energy approximations of Brane World gravity do not hold up for this potential. Thus for inverse monomial inflation we see that 
the value of $p$ must be greater than 2 if we consider the high energy expressions of brane-world.

Now we look at specific cases of the potential for given values of p :

\subsubsection{p=3 : $V(\phi=M^4\ppl^{-3}$}
\begin{itemize}
 \item {{\bf{Spectral Index}}} ($n_s)= 0.8347$
 \item{{\bf{Running of the spectral index}}} ($\alpha_s)= -0.01748$
 \item{\bf{{Tensor to scalar ratio}}} ($r) =5.79\times 10^{-29}$
\end{itemize}

\subsubsection{p=4 : $V(\phi=M^4\ppl^{-4}$}

\begin{itemize}
 \item {{\bf{Spectral Index}}} ($n_s)= 0.8833$
 \item{\bf{Running of the spectral index}} ($\alpha_s)= -8.611\times 10^{-3}$
 \item{\bf{Tensor to scalar ratio}} ($r)=6.28\times 10^{-29}$
\end{itemize}

\subsubsection{p=5 : $V(\phi=M^4\ppl^{-5}$}
\begin{itemize}
 \item {{\bf{Spectral Index}}} ($n_s)= 0.8994$
 \item{{\bf{Running of the spectral index}}} ($\alpha_s)= -6.4026\times 10^{-3}$
 \item{\bf{{Tensor to scalar ratio}}} ($r)=7.10\times 10^{-29}$
\end{itemize}

\subsubsection{p=6 : $V(\phi=M^4\ppl^{-6}$}
\begin{itemize}
 \item {{\bf{Spectral Index}}} ($n_s)= 0.9075$
 \item{{\bf{Running of the spectral index}}} ($\alpha_s)= -5.3667\times 10^{-3}$
 \item{\bf{{Tensor to scalar ratio}}} ($r)=7.25\times 10^{-29}$
\end{itemize}

\subsection{Open String Tachyonic Inflation(OSTI): $V(\phi)=-M^4 \left(\frac{\phi}{\phi_0}\right)^2
\ln{\left[\left(\frac{\phi}{\phi_0}\right)\right]^2}$}

We first let $x=\frac{\phi}{\phi_0}$ and therefore $V(x)=-M^4 x^2 \ln(x^2)$. 

\begin{figure}[h!!!!]
\centering
\epsfig{figure=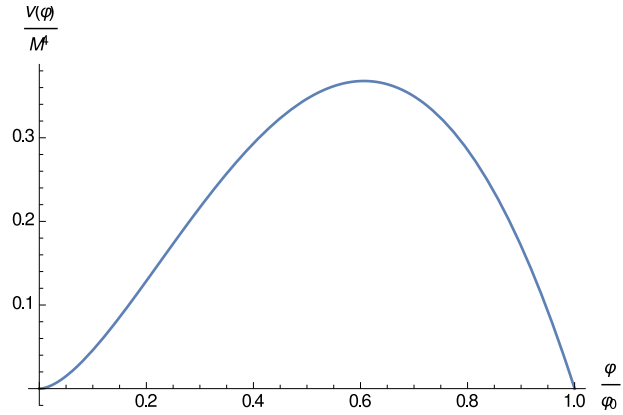,width=8cm,angle=0}
\caption{{\bf OSTI} : Slow roll inflation proceeds downhill from left to right in the above figure. The value of $M^4$ is given 
as $M^4=e\lambda$ and $\phi_0=20 M_4$.}
\label{osti}
\end{figure}

The derivatives of the potential are :
\begin{eqnarray}
\label{deosti}
V'&=&\frac{dV}{d\phi}=\frac{dV}{dx}\frac{dx}{d\phi}= -\frac{2M^4}{\phi_0}{x\left[1+\ln(x^2)\right]} \nonumber \\ 
V''&=& \frac{d^2V}{d\phi^2}=\frac{d}{dx}\left(\frac{dV}{d\phi}\right)\frac{dx}{d\phi}=-\frac{2M^4}{\phi_0^2}(3+\ln(x^2))
\nonumber \\ V'''&=&\frac{d}{d\phi}\left(\frac{d^2V}{d\phi^2}\right)= -\frac{4M^4 x}{\phi_0^3}
\end{eqnarray}

The slow roll parameters for this potential are from equations (\ref{hee}, \ref{heeta}, \ref{hexi}, \ref{deosti}) :
\begin{eqnarray}
\label{ostie}
\epsilon_v\approx \frac{\lambda M_4^2}{4\pi}\frac{V'^2}{V^3}  \approx -\frac{\lambda M_4^2}{\pi \phi_0^2 M^4}\left[\frac{(1+ \ln(x^2))^2}{x^4 (\ln (x^2))^3}\right]
\end{eqnarray}

and
\begin{eqnarray}
\label{ostieta}
\eta_v\approx \frac{\lambda M_4^2}{4\pi}\left(\frac{V''}{V^2}\right) \approx-\frac{\lambda M_4^2}{2\pi\phi_0^2 M^4}
\left[{3+\ln(x^2)}{x^4 \ln(x^2)}\right]
\end{eqnarray}
and
\begin{eqnarray}
\label{ostixi}
\xi_v^2\approx \frac{\lambda^2 M_4^2}{16\pi^2}\frac{V'V'''}{V^4} \approx \frac{\lambda^2 M_4^2}{16\pi^2}
\left[-\frac{2M^4}{\phi_0^2}\times -\frac{4M^4}{\phi_0^2}\times \frac{x(3+\ln(x^2))(1+\ln(x^2)}{M^{16}x^8(\ln(x^2))^4}\right]
\end{eqnarray}

The No. of e-folds are :
\begin{eqnarray}
N&\approx& -\frac{4\pi}{\lambda M_4^2}\int_{\phi_{i}}^{\phi_{end}} \frac{V^2}{V'} d\phi \approx \frac{2\pi M^4 \phi_0^2}{\lambda M_4^2}\int_{x_{i}}^{x_{end}} \frac{x^3 (\ln(x^2))}{1+\ln(x^2)} dx
\nonumber \\ &\approx& \frac{2\pi M^4 \phi_0^2}{\lambda M_4^2}\left[-\frac{3}{8} x^4 +\frac{1}{4} x^4 \ln(x^2)\right]_{x_{i}}
^{x_{end}} + \nonumber \\ &&\frac{\pi M^4 \phi_0^2}{\lambda M_4^2 e^2} ExpIntegralEi\left[2(1+\ln(x^2))\right]_{x{i}}^{x_{end}}
\nonumber \\ &\approx& \frac{2\pi M^4 \phi_0^2}{\lambda M_4^2}\left[-\frac{3}{8} x_{end}^4 + \frac{1}{4} x_{end}^4\ln(x_{end}^2)
  +\frac{3}{8} x_{i}^4 -\frac{1}{4} x_{i}^4 \ln(x_{i})^2\right] \nonumber \\ && 
  -\frac{\pi M^4 \phi_0^2}{\lambda M_4^2 e^2} ExpIntegralEi\left[2(1+\ln(x_{i}^2))\right] \nonumber \\ &&+
  \frac{\pi M^4 \phi_0^2}{\lambda M_4^2 e^2} ExpIntegralEi\left[2(1+\ln(x_{end}^2))\right]
\end{eqnarray}
where $ExpIntegralEi$ is the Exponential Integral function \cite{hb}  $Ei(z)= -\int_{-z}^\infty \frac{exp{(-t)}}{t} dt$
 where the principal value of the integral is taken.

From High Energy considerations we have :
\beq
\label{ostihi}
M^4= e\lambda
\eeq
 The form of the amplitude of the scalar power spectrum is :
\begin{eqnarray}
A_s^2\approx \frac{64}{75}\frac{\pi}{\lambda^3 M_4^6}\left(\frac{V^6}{V'^2}\right) \approx\frac{16\pi\phi_0^2}{75 \lambda^3 M_4^6} M^{16}\frac{x_{*}^{10}(\ln(x_{*}^2))^6}
{(1+\ln(x_{*}^2))^2}
\label{ostis}
\end{eqnarray}
where $x_{*}$ is the value of the field variable at horizon crossing $k=aH$.

Now putting the value of $M^4$ (\ref{ostihi}) in equation (\ref{ostis}) we get :
\begin{eqnarray}
\left(\frac{1+\ln(x_{*}^2)}{x^5 (\ln(x_{*}^2))^3}\right)^2\approx 1.6409\times 10^{-78}\left(\frac{\phi}{M_4}\right)^2
\end{eqnarray}
\\
So we see that we can calculate for $x_{*}$ if we knew $\phi_0$. So we can find a range of values of $x_{*}$ for 
different $\phi_0$. Assuming $\phi_0=20 M_4$ we get:
\beq
\left(\frac{1+\ln(x_{*}^2)}{x_{*}^5 (\ln(x_{*}^2))^3}\right)\approx 1.2809\times 10^{-39}
\eeq

Now we plot this function graphically and get the value of $x_{*}$ from it (Fig: \ref{opst}).
\begin{figure}[h!!!!!!!!]
\centering
\epsfig{figure=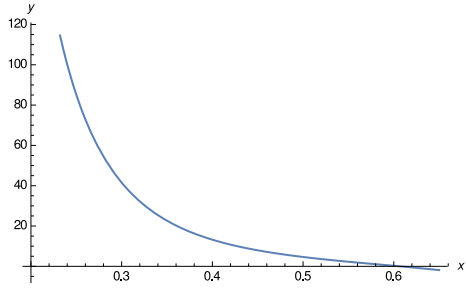,width=6cm,angle=0}
\caption{Plot of $y= \left(\frac{1+\ln(x_{*}^2)}{x_{*}^5 (\ln(x_{*}^2))^3}\right)$ vs $x_{*}$ to get value of $x_{*}$.}
\label{opst}
\end{figure}
\\

We find the  value of $x_{*}$ as :
\beq
\label{ostic}
x_{*}=0.6065
\eeq

The values of the slow roll parameters at horizon crossing is from equations (\ref{ostie}, \ref{ostieta}, \ref{ostixi}, \ref{ostic}) :
\begin{eqnarray}
\epsilon_v= 2.1629\times 10^{-4} ,
\eta_v= -2.1612\times 10^{-3} ,
\xi_v= -2.8688\times 10^{-10}
\end{eqnarray}

Thus the values of $n_s$ and $\alpha_s$ from equation(\ref{hens},\ref{heas}) are :
\begin{eqnarray}
n_s&=&\boxed{0.9957}  \\ \alpha_s&=&\boxed{8.26\times 10^{-6}}
\end{eqnarray}

The tensor to scalar ratio is :
\begin{eqnarray}
r\approx \epsilon_v\frac{\lambda}{V}\approx \boxed{2.16\times 10^{-11}} 
\end{eqnarray}

\section{Summary of Results}

We summarise in tabular form the results of $n_s,\alpha_s$ and $r$ for the various models that we have considered. For the sake 
of comparison we include the results of Planck \cite{Pl} and WMAP \cite{Wm}.\\\\
We reproduce the table for Planck and WMAP : \\\\
\begin{tabular}{|c|c|c|}
 \hline
 Observational Parameters & Planck & WMAP (9 year)\\
 \hline
 $r$(Tensor-to-scalar ratio) & $<0.11$ & $<0.13$\\
 \hline
 $n_s$(scalar spectral index) & $0.968\pm0.006$ & $0.972\pm0.013$\\
 \hline
 $\vartriangle_s^2$ & $(2.23\pm0.16)\times10^{-9}$ & $(2.41\pm0.10)\times10^{-9}$\\
 \hline
 $\alpha_s$ & $(-0.003\pm0.007)$ & $(-0.019\pm0.025)$\\
 \hline
\end{tabular}\\\\

\begin{tabular}{|c|c|c|c|}
\hline
Inflationary Potentials ($V(\phi)$)& ($n_s$) & ($\alpha_s$) & ($r$)\\
\hline
Large Field Inflation& & &
\\p=1 &0.9666 &-7.85$\times 10^{-4}$ &9.9767$\times 10^{-27}$\\
p=2/3 &0.9699&-5.9028$\times 10^{-4}$& 8.24$\times10^{-27}$\\
p=2&0.9572&-1.1722$\times 10^{-3}$&4.31$\times10^{-26}$\\
p=3&0.9522 &-4.3098$\times 10^{-3}$ &4.86$\times 10^{-26}$\\
p=4&0.9489 &-7.6030$\times 10^{-3}$ &5.22$\times 10^{-26}$\\
p=5&0.9465 &-10.5314$\times 10^{-3}$ &5.51$\times 10^{-26}$\\
p=6&0.9447 &-13.0557$\times 10^{-3}$ &5.62$\times 10^{-26}$\\
\hline
Power Law Inflation&0.9320& -2.8866$\times 10^{-3}$& 1.3807$\times 10^{-28}$\\
\hline
Inverse Monomial Inflation& & &\\p=3 & 0.8347 & -17.48$\times 10^{-3}$ & $5.79\times 10^{-29}$\\
p=4 &0.8833&-8.611$\times 10^{-3}$&$6.28\times 10^{-29}$ \\
p=5 &0.8999 &-6.4026$\times 10^{-3}$ &$7.10\times 10^{-29}$\\
p=6 &0.9075 &-5.3667$\times 10^{-3}$ &$7.25\times 10^{-29}$\\
\hline
Open String Tachyonic Inflation & 0.9957 &8.26$\times 10^{-6}$&$2.16\times 10^{-11}$ \\
\hline
\end{tabular}
\\\\
We can make some general comments from this results.
\begin{itemize}
 \item{\bf{Low value of $r$}} : We can see that the Brane World results for the tensor to scalar ratio for all the models discussed 
 are very low and beyond the reach of experimental probes even in the foreseeable future. Thus an important parameter for 
 G.R. based perturbative analysis is already ruled out as an experimentally testable quantity in Brane-World. However if we 
 can constrain the values of $r$ like the BICEP2 \cite{bicep} results then we can comprehensively rule out all the existing models 
 in the context of Brane-World Gravity. This would bring us to question the underlying theory: Is Brane- World gravity an 
 accurate physical theory? Clearly all existing models of inflation cannot be wrong so there must be some problems with 
 the original theory as well. That will be something interesting to look out for. It also needs to be pointed out that such 
 extremely low values of $r$ deprives us of an important parameter for ruling in or ruling out inflationary models as given 
the current upper bound on $r$ from experimental results they are allowed even if they are impossible to be found experimentally.
\item{\bf{Values of $\alpha_s$ and $n_s$ not simultaneously conforming to experiment}}: 
The values of $\alpha_s$ are beyond the Planck cut-off limit for most 
of the models which return a suitable (experimentally allowed) $n_s$ in Brane-World gravity. Again the models for which the 
values of $\alpha_s$ are experimentally allowed fall foul of the similar predictions for $n_s$. 
\item{\bf Clustering of $\alpha_s$ around a certain range} : There seems to be a 
clustering of the values around a specified range for $\alpha_s$ around the $10^{-3}$ range with only OSTI, LFI$_1$ and LFI$_{2/3}$ 
falling outside that range. A similar clustering was seen in a G.R. based analysis in \cite{univ}. Intriguingly the same holds 
true in Brane World gravity.
\item{\bf{Values of $n_s$ shifts away from 1}}: We had seen that theoretically the expectation was that the spectral index should be 
pushed towards a Harrison Zeldovich spectrum. However as we can see that except OSTI all the models have spectral indices 
pushed away from 1 even more than the G.R. case. This was reported for just the $LFI_2$ potential in \cite{chbr} but it is seen 
to be a general trend for the models in brane-world.
This is a significant break from what we expect theoretically.
  \item{\bf{Large Field Inflation}} : Chaotic inflation is ruled  out we consider the value for the spectral index 
and the running of the spectral index for the Planck 2015 \cite{Pl} data. However it was ruled in for the Planck 2013 \cite{Pl1}. 
The shift towards higher $n_s$ values for Planck 2015 results compared to Planck 2013 data leads to the $p=2$ model being ruled out.
However for the values of $p=4, 5, 6$ the potential is ruled out if we consider $n_s$ while giving acceptable 
values for $\alpha_s$ illustrating the point that we made previously. The values of the potential for $p=1, 2/3$ are 
ruled in by the data. The $p=3$ model is ruled out by the Planck (2015) and WMAP (9 year) results. It was previously ruled in by 
the Planck (2013) results.
\item{\bf{Power Law Inflation}} : One idiosyncrasy of this model in Brane World gravity is that the values of $M^4$ is not constrained by the scalar 
amplitude but the value of $\alpha$ is. However that is not a problem in our calculations where the quantities can always be expressed 
in terms of $\alpha$. The model is ruled out in Brane World gravity for both $n_s$ and $\alpha_s$.
\item{\bf{Inverse Monomial Inflation}} : This model only works in Brane World gravity for $p>2$ as is also seen by a slightly different 
analysis in \cite{eujr}. However no definite reason is known for this. The high energy corrections of brane-world obviously 
do not hold up for $p<2$. Interestingly this model with $p=1,2,3,4,5,6$ gives physical results in a G.R. based analysis.
For $p=3,4,5,6$ the Brane World corrections are significant. The results are however incompatible with 
experimental results. As $p$ increases the models come closer to the lower values of $n_s$ from 
the Planck and WMAP results while the $\alpha_s$ values are generally allowed by the data. 
\item{\bf{OSTI}} : This model is ruled out by the present data. While the value of $n_s$ is pushed towards 1 which while agreeing 
with theoretical results is a significant break from what we expect experimentally and what we get for the other models. The value of 
$\alpha_s$ is also beyond the cutoff of the Planck and WMAP data.
\end{itemize}

\chapter{Conclusions}

In this report we have worked out the cosmological observables for four distinct classes of potentials and checked their results 
with the available experimental data. While more work needs to be done with other models of inflation potentials before anything 
definitive can be proclaimed we see that for the models considered the theoretical results are in tension with the 
experimental data. None of the models considered simultaneously fit the data for $\alpha_s$ and $n_s$. However a dedicated 
effort to study the well favoured models (from a G.R. perspective) after Planck (2015) from a Brane World context is essential 
before we can make definite predictions. Importantly the presence of an unconstrained $M^4$ term in the slow roll parameters 
in the high energy expressions of the slow roll parameters is the root of all trouble and the reason why more models could 
not be tackled adequately. Instructively the G.R. based theory is free from this malaise. 

The cosmological observables also show distinct trends with the tensor to scalar ratio tending to be very small, the spectral index 
shifting away from 1 (contrary to theoretical predictions) and the running of the spectral index clustering around a particular value 
of the order of $10^{-3}$.  Interestingly a similar 
clustering of $\alpha_s$ is seen in a G.R. based analysis. We plan to study more models to see whether these are in general predictions 
and with better and better results coming of the CMB power spectrum this analysis will become useful in the near future 
for acting as a falsifiability check on Brane World gravity in the context of inflationary dynamics.

\newpage
\vspace{1cm}

\begin{center}

{\Large \bf{Acknowledgements}}
\end{center}
\vspace{1cm}

I would like to thank my advisor Dr. Ratna Koley for her guidance and help all throughout these four months, for nudging me 
in the right direction every time I strayed and for going through all the reports and painstakingly pointing out all the errors and 
mistakes. I would also like to thank all my friends and my family for supporting me and picking me up whenever I was down. 
I also thank the academic and non-academic members of the Physics Department and I am 
indebted to the facilities of the Baker Lab which I used throughout this semester. I would
also like to thank the online article repositories like arXiv and packages like Mathematica 
which were a big help for this project. Finally I would like to 
thank the Planck and WMAP missions for the data of the CMB which was the cornerstone of this project.

\end{document}